\documentclass{SciPost}

\usepackage{graphicx}
\usepackage{epstopdf}

\usepackage{booktabs}
\usepackage{dcolumn}
\newcolumntype{d}[1]{D{.}{.}{#1}}
\usepackage{longtable}

\hypersetup{colorlinks,citecolor=green!40!black,urlcolor=blue!40!black,linkcolor=red!90!black}
\usepackage{cite,mcite}

\gdef\ffrac#1#2{\textstyle\frac{#1}{#2}\displaystyle}

\begin{document}

\begin{center}{\Large \textbf{
Breathing distortions in the metallic, antiferromagnetic phase of
LaNiO$_3$ }}\end{center}

\begin{center}
Alaska Subedi\textsuperscript{1,}\textsuperscript{2}
\end{center}

\begin{center}
{\bf 1} Centre de Physique Th\'eorique, \'Ecole Polytechnique, CNRS,
  Universit\'e Paris-Saclay, F-91128 Palaiseau, France
\\
{\bf 2} Coll\`ege de France, 11 place Marcelin Berthelot, 75005
  Paris, France
\\
\end{center}

\begin{center}
\today
\end{center}


\section*{Abstract}
{\bf I study the structural and magnetic instabilities in LaNiO$_3$
  using density functional theory calculations. From the
  non-spin-polarized structural relaxations, I find that several
  structures with different Glazer tilts lie close in energy. The
  $Pnma$ structure is marginally favored compared to the
  $R\overline{3}c$ structure in my calculations, suggesting the
  presence of finite-temperature structural fluctuations and a
  possible proximity to a structural quantum critical point. In the
  spin-polarized relaxations, both structures exhibit the
  $\uparrow\!\!0\!\!\downarrow\!\!0$ antiferromagnetic ordering with a
  rock-salt arrangement of the octahedral breathing distortions. The
  energy gain due to the breathing distortions is larger than that due
  to the antiferromagnetic ordering. These phases are semimetallic
  with small three-dimensional Fermi pockets, which is largely
  consistent with the recent observation of the coexistence of
  antiferromagnetism and metallicity in LaNiO$_3$ single crystals by
  Guo \textit{et al.}  [Nat.\ Commun.\ {\bf 9}, 43 (2018)].}

\vspace{10pt}
\noindent\rule{\textwidth}{1pt}
\tableofcontents\thispagestyle{fancy}
\noindent\rule{\textwidth}{1pt}
\vspace{10pt}

\section{Introduction}
\label{sec:intro}

The rare-earth nickelates $R$NiO$_3$ have received an enduring
interest over the last two and half decades because they exhibit
unique structural and electronic transitions that can be tuned from
600 to 130 K as a function of the rare-earth ion $R$
\cite{Medarde1997,Catalan2008,Balents2017}. All rare-earth nickelates
except LaNiO$_3$ \cite{Goodenough1965} occur in a perovskite-type
orthorhombic structure with the space group $Pnma$ in the
high-temperature phase, which is metallic \cite{Torrance1992}. As the
temperature is lowered, all orthorhombic rare-earth nickelates undergo
a structural transition to a monoclinic structure with the space group
$P2_1/n$ \cite{Alonso1999a,Alonso1999b,Alonso2000,Alonso2001}. This
structural transition involves a splitting of the Ni positions into
two inequivalent sites, and it simultaneously transforms these
materials into insulators. Additionally, these materials also undergo
a magnetic transition that coincides with the structural and
electronic transition temperature for $R =$ Pr and Nd, but occurs at a
lower temperature for the rare-earth ions with smaller radii.

Although the electronic and magnetic transitions in the orthorhombic
rare-earth nickelates show relatively large responses in the
resistivity and susceptibility measurments, identifying the order
parameters and the microscopic mechanism behind these transitions has
remained challenging. The metal-insulator and
paramagnetic-antiferromagnetic transitions in these materials were
observed as early as 1989 \cite{Torrance1992,Vassiliou1989}. However,
the monoclinic $P2_1/n$ structure of the low-temperature phase was
only resolved in 1999 \cite{Alonso1999a,Alonso1999b}. The magnetic
ordering occurs at the wave vector $(\ffrac12,\ffrac12,0)_o$ relative
to the orthorhombic unit cell
\cite{Garcia1992,Garcia1994,Rodriguez1998}. But several arrangements
of the magnetic moments are consistent with the available neutron
scattering data, and the magnetic structure of the ordered phase has
not been fully determined.

The metal-insulator transition in the orthorhombic rare-earth
nickelates was initially believed to be a Mott transition
\cite{Torrance1992}. However, the observation of the compression and
expansion of the NiO$_6$ octahedra in an alternating manner in the
low-temperature monoclinic phase makes this explanation untenable and,
instead, points towards a charge ordering mechanism
\cite{Alonso1999a}. The Ni$^{3+}$ ions have a nominal occupancy of
$e_{g}^1$ in these materials, and the absence of Jahn-Teller
distortion for this electronic configuration is also
surprising. Goodenough and Raccah have argued that this absence is due
to a large covalency between the Ni $3d$ and O $2p$ orbitals, which
makes the antibonding $e_{g}^1$ state highly
delocalized\cite{Goodenough1965,Goodenough1996}. Hartree-Fock cluster
calculations by Mizokawa \textit{et al.}\ that took into account the
large Ni $3d$--O $2p$ covalency found breathing distortions of the O
ions to be stable for the small rare-earth ion nickelates
\cite{Mizokawa2000}. However, for larger rare-earth nickelates such as
PrNiO$_3$ and NdNiO$_3$, they found the displacement of the Ni ions
along the cubic diagonal direction to be favorable, which was not
observed in high-resolution diffraction experiments
\cite{Medarde2008,Garcia2009}.

The presence of a large Ni $3d$--O $2p$ covalency may explain the lack
of Jahn-Teller distortion in the rare-earth nickelates, but this
concept does not identify the microscopic instability that causes the
breathing distortions in all the rare-earth nickelates except
LaNiO$_3$. A major leap in understanding the phase transitions in the
rare-earth niceklates was achieved by the insight of Mazin \textit{et
  al.}, who showed that a charge ordering of the type $2e_g^1
\rightarrow e_g^0 + e_g^2$ occurs when the on-site Hund's rule
coupling $J$ overcomes the on-site Coulomb repulsion (i.e., when $U -
3J$ is small) \cite{Mazin2007}. This charge ordering of the
antibonding $e_g$ electrons naturally leads to an expansion and
compression of the alternate NiO$_6$ octahedra that is experimentally
observed in the low-temperature monoclinic phase of the rare-earth
nickelates. Additionally, it also gives rise to the magnetic ordering
$\uparrow\!\!0\!\!\downarrow\!\!0$ (antiferromagnetic ordering of the
Ni moments in the larger octahedra and absence of moments at the Ni
sites in the smaller octahedra), which is consistent with the
available neutron scattering data.

The mechanism for the phase transitions in the rare-earth nickelates
suggested by Mazin \textit{et al.}'s phenomenological model and
density functional theory (DFT) calculations has been further
supported by dynamical mean field theory (DMFT) calculations using
realistic electronic structures that span both the high-energy Ni $3d$
$+$ O $2p$ \cite{Park2012} and low-energy antibonding $e_g$ energy
scales \cite{Subedi2015}. Other theoretical studies using diverse
techniques have also supported this description of the phase
transition \cite{Johnston2014,Green2016,Varignon2017,Lu2017}. The DMFT
calculations utilizing the low-energy $e_g$ manifold further
highlighted the essential role played by the breathing distortions in
causing the metal-insulator transition \cite{Subedi2015}. It was found
that even small breathing distortions splits the quarter filled $e_g$
bands, resulting in a system with a manifold of narrow half-filled
bands. Alternatively, in the real space picture, the distortion makes
the on-site energies of the neighboring quarter-filled $e_g$ states
inequivalent, which causes one site to be half-filled and another to
be empty. This change in the electronic structure, viewed from either
picture, makes the system highly susceptible to undergo a transition
to an insulating phase.

Moving the focus to the title compound LaNiO$_3$, it is curious that
it shows a behavior that is distinct from all other rare-earth
nickelates even though the ionic radius of La$^{3+}$ is close to that
of the early members of the lanthanide series such as
Pr$^{3+}$. LaNiO$_3$ occurs in the rhombohedral $R\overline{3}c$
structure and is not known to exhibit any structural or
metal-insulator transitions, unlike other rare-earth
nickelates. However, several experiments have hinted at the proximity
of LaNiO$_3$ to other rare-earth nickelates. The high-temperature
magnetic susceptibility, resistivity, and thermoelectric power of both
LaNiO$_3$ and the orthorhombic rare-earth nickelates display similar
features that are consistent with the presence of a heterogeneous
phase consisting of two different Ni sites
\cite{Zhou2000a,Zhou2000b,Zhou2004}. Optical and electron tunneling
spectroscopy experiments show the presence of a pseudogap in thin
films of both LaNiO$_3$ and NdNiO$_3$ \cite{Stewart2011a,Allen2015}. A
pair density function analysis of the neutron scattering data of a
powder LaNiO$_3$ sample by Li \textit{et al.} found that the local
structure is better described by monoclinic $P2_1/n$ and orthorhombic
$Pnma$ structures below and above 200 K, respectively \cite{Li2016}.

The lack of high-quality single crystals has hindered a more rigorous
determination of the properties of the rare-earth
nickelates. Recently, two groups have reported the synthesis of
LaNiO$_3$ single crystals using the floating zone technique under high
oxygen pressures. Zhang \textit{et al.}'s samples, which were grown
under the oxygen pressure of 30--50 bar, were characterized to have
the rhombohedral $R\overline{3}c$ structure and showed metallic
conductivity \cite{Zhang2017}. These samples did not exhibit any
structural or magnetic transition, but their magnetic susceptibility
showed a broad maximum around 200 K. A small anomaly had also been
observed around this temperature in earlier measurements of the
magnetic susceptibility on polycrystalline samples \cite{Li2016}.
However, the samples grown by Guo \textit{et al.}\ under the oxygen
pressure of 130--150 bar showed an antiferromagnetic transition at 157
K with an ordering wave vector of $(\ffrac14,\ffrac14,\ffrac14)_c$ in
the pseudocubic notation \cite{Guo2018}. The antiferromagnetic phase
remained in the rhombohedral $R\overline{3}c$ structure and continued
to exhibit a metallic behavior. This is rather surprising considering
that the pair density function analysis mentioned above indicated a
structural similarity at the nanoscale between LaNiO$_3$ and other
rare-earth nickelates \cite{Li2016}. If an antiferromagnetic
transition were to be present in LaNiO$_3$, one would have expected it
to also show structural and metal-insulator transitions like other
rare-earth nickelates.

The presence or absence of the breathing distortions in the
antiferromagnetic phase of LaNiO$_3$ has important implications on the
microscopic mechanism for the phase transition in the rare-earth
nickelates. Lee \textit{et al.} have suggested that Fermi surface
nesting, not charge disproportionation, plays a key role in the phase
transition of the rare-earth nickelates, especially the ones with
larger rare-earth radii \cite{Lee2011a,Lee2011b}. They have shown that
charge disproportionation necessarily occurs in the orthorhombic
$Pnma$ rare-earth nickelates as a secondary order parameter during the
antiferromagnetic phase transition. On the other hand, their symmetry
analysis within a Landau theory suggested that a pure
antiferromagnetic state without any disproportionation occurs in the
rhombohedral $R\overline{3}c$ phase. A lack of breathing distortions
in antiferromagnetic LaNiO$_3$ would imply that the disproportionation
suggested by Mazin \textit{et al.}\ does not play a decisive role.

The above discussion amply demonstrates that the rhombohedral
LaNiO$_3$ is close to the structural and magnetic phases that appear
in other rare-earth nickelates. DFT calculations and its extensions
DFT+$U$ and DFT+DMFT have been used to study the structural,
electronic, and magnetic properties of the rare-earth nickelates
\cite{Hamada1993,Mazin2007,Giovannetti2009,May2010,Chakhalian2011,
  Gou2011,Deng2012,Park2012,Parragh13,Park2014,Peil2014,Subedi2015,Nowadnick2015,
  Varignon2017,Hampel2017,Haule2017,Seth2017}. Park \textit{et
  al.}\ have claimed that DFT is inadequate to qualitatively describe
the ground-state properties of the rare-earth nickelates and more
sophisticated methods are necessary \cite{Park2014}.  However, recent
works by Varignon \textit{et al.}\ \cite{Varignon2017} and Hampel and
Ederer \cite{Hampel2017} show that rigorous DFT calculations can
describe the antiferromagnetic and disproportionated ground state of
these materials. These authors focused their studies on the
orthorhombic rare-earth nickelates. A similar study on LaNiO$_3$ would
be helpful in clarifying the structural, electronic, and magnetic
properties of this material.

In this paper, I use DFT calculations to explore the structural,
electronic, and magnetic instabilities of LaNiO$_3$. The calculated
non-spin-polarized phonon dispersions of cubic LaNiO$_3$ show
instabilities at the wave vectors $R$ $(\ffrac12,\ffrac12,\ffrac12)_c$
and $M$ $(\ffrac12,\ffrac12,0)_c$. I fully relaxed the supercells that
exhibit the various Glazer tilts allowed by these instabilities. I
find that several structures with different Glazer tilts lie close in
energy. The orthorhombic $Pnma$ structure is marginally lower in
energy than the rhombohedral $R\overline{3}c$ structure in my
calculations. This suggests the presence of structural fluctuations in
LaNiO$_3$ at finite temperatures and indicates a possible proximity to
a structural quantum critical point. Both structures exhibit the
$\uparrow\!\!0\!\!\downarrow\!\!0$ antiferromagnetic order with a
rock-salt arrangement of the octahedral breathing distortions when the
spin-polarized relaxations are performed. The gain in energy due to
the breathing distortions is larger than the gain in energy due to the
antiferromagnetic ordering, suggesting that the mechanism of
disproportionation proposed by Mazin \textit{et al.}\ plays a key role
in the phase transition of LaNiO$_3$. These phases are semimetallic
with small three-dimensional Fermi pockets. This is mostly consistent
with the recent experiments of Guo \textit{et al.}\ that uncovered an
antiferromagnetic transition in LaNiO$_3$ without a concomitant
metal-insulator transition \cite{Guo2018}. They did not observe the
breathing distortions that I find in my calculations, and this might
be because the calculated difference of $\sim$0.01 \AA\ between the
Ni-O bond lengths of the compressed and expanded octahedra is very
small.

\section{Methods}
\label{sec:meth}

The DFT calculations presented here were obtained using the
pseudopotential-based planewave method as implemented in the {\sc
  quantum espresso} package \cite{qe}. The phonon dispersions were
calculated using density functional perturbation theory
\cite{dfpt}. The calculations were done within the generalized
gradient approximation of Perdew, Burke and Ernzerhof (PBE
GGA)\cite{pbe} using the pseudopotentials generated by Garrity
\textit{et al.}\ \cite{gbrv}. Some calculations were also checked
using the ONCV pseudopotentials \cite{oncv}, as well as Garrity
\textit{et al.}'s pseudopotentials within the local density
approximation (LDA). The planewave basis-set and charge density
expansions were done using cut-offs of 50 and 250 Ry, respectively.

I used a $16\times16\times16$ $k$-point mesh for the Brillouin zone
integration in the phonon calculations.  The dynamical matrices were
obtained on an $8\times8\times8$ $q$-point grid which includes the
special high-symmetry points $R$ $(\frac12,\frac12,\frac12)_c$ and $M$
$(\frac12,\frac12,0)_c$ emphasized below. The phonon dispersions were
obtained by Fourier interpolation. The structural relaxation of the
various Glazer tilts \cite{Glazer} were done on 40-atom
$2\times2\times2$ pseudocubic supercells using an $8\times8\times8$
$k$-point mesh. I used denser meshes in the spin-polarized structural
relaxations. For the 20- and 80-atom supercells of the
$R\overline{3}c$ structure, I used $12\times12\times8$ and
$8\times8\times8$ meshes, respectively. For the 40- and 80-atom
supercells of the $Pnma$ structure, $6\times8\times12$ and
$6\times4\times12$ meshes were used, respectively.  I made extensive
use of the {\sc isotropy} \cite{isotropy} and {\sc spglib}
\cite{spglib} packages in the symmetry analysis. {\sc vesta}
\cite{vesta} and {\sc xcrysden} \cite{xcrysden} were used to visualize
the crystal structures and Fermi surfaces, respectively.

To check convergence with respect to the planewave cut-off, I repeated
the structural relaxations of the various Glazer tilts for a cut-off
value of 60 Ry. Same energetic rankings were obtained. I also did some
calculations with larger $k$-point meshes, and this did not change the
results in a meaningful way. Note that the meshes used in this work
are denser than those used in two recent DFT studies on the rare-earth
niclelates \cite{Varignon2017,Hampel2017}.

\section{Results and Discussions}
\label{sec:resu}

\subsection{Non-spin-polarized structural relaxations}

\begin{figure}
  \includegraphics[width=0.5\textwidth]{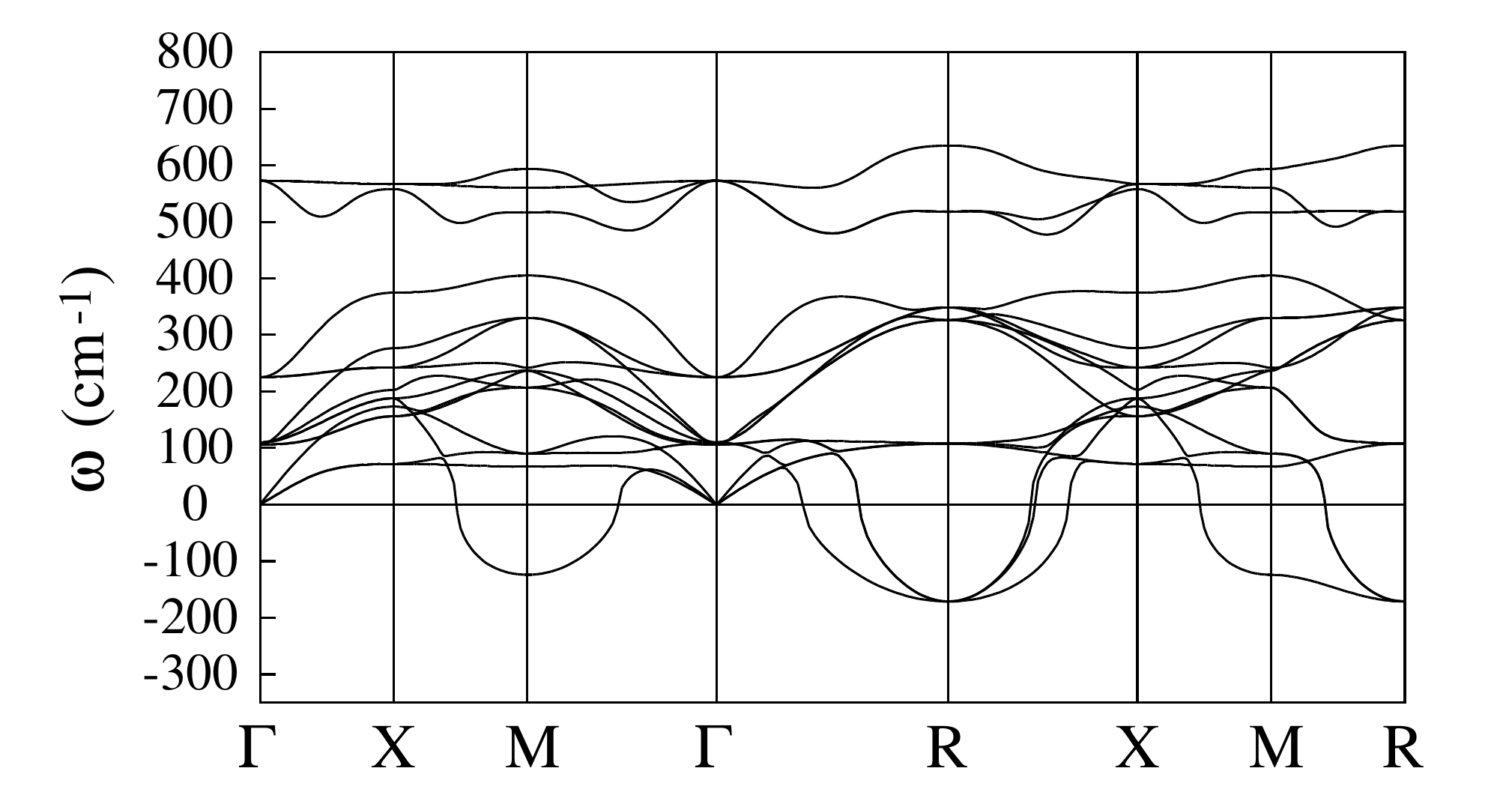}
  \includegraphics[width=0.5\textwidth]{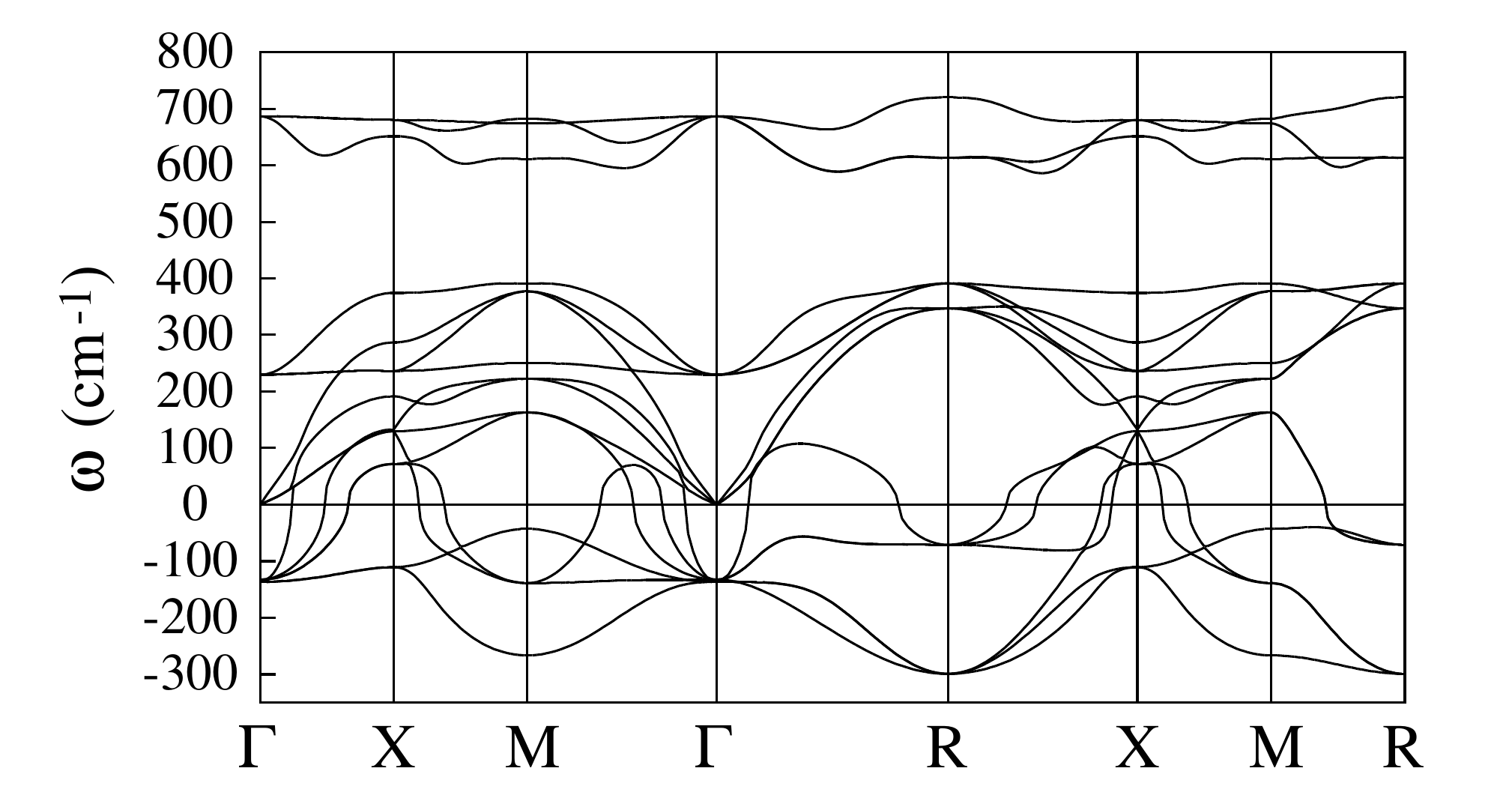}
  \caption{Calculated PBE GGA non-spin-polarized phonon dispersions of
    cubic LaNiO$_3$ (left) and cubic YNiO$_3$ (right). The imaginary
    frequencies are denoted by negative values.}
  \label{fig:cpband}
\end{figure}

The rhombohedral $R\overline{3}c$ structure of LaNiO$_3$ is
characterized by out-of-phase rotations of the oxygen octahedra about
the three axes of the parent cubic phase and is denoted by $a^-a^-a^-$
in Glazer's notation. The orthorhombic $Pnma$ structure of all other
rare-earth nickelates involves an in-phase rotation about one axis and
out-of-phase rotations by a different amount about the two other axes
and has the notation $a^+b^-b^-$.  These distorted structures derive
from different dynamical instabilities of the cubic perovskite
phase. To examine if they have similar latent structural
instabilities, I start by comparing the non-spin-polarized phonon
dispersions of cubic LaNiO$_3$ and, as a representative member of the
orthorhombic family, cubic YNiO$_3$, which are shown in
Figs.~\ref{fig:cpband}(a) and (b), respectively. These were calculated
within the PBE GGA using the relaxed lattice parameters of $a$ = 3.837
and 3.745 \AA\ for LaNiO$_3$ and YNiO$_3$, respectively.

The phonon dispersions of the two nickelates in the cubic phase show
several branches that are unstable along different directions in the
Brillouin zone. The phonon instabilities are weaker in LaNiO$_3$ than
in YNiO$_3$, consistent with the observation that LaNiO$_3$ is closer
to the cubic phase. In both materials, the largest instability occurs
at the wave vector $R$ $(\ffrac12,\ffrac12,\ffrac12)_c$ in the
pseudocubic notation. This mode is triply degenerate and has the
irreducible representation (irrep) $R^+_4$ when the convention that Ni
is placed at $(0,0,0)$ is used. [Its irrep is $R^-_5$ when Ni is at
  $(0.5,0.5,0.5)$.] Another mode at $M$ $(\ffrac12,\ffrac12,0)_c$ also
shows a large instability. It is singly degenerate and has the irrep
$M^+_3$ when Ni is placed at $(0,0,0)$. The $R^+_4$ mode induces
out-of-phase rotations of the oxygen octahedra about the axes, while
the $M^+_3$ mode generates in-phase rotations. Note that although the
unstable phonon at $M$ is singly degenerate, the star of $M$ has three
points. So the $M^+_3$ mode is able to generate in-phase rotations
about all three axes.

The $a^-a^-a^-$ tilt system of the rhombohedral LaNiO$_3$ requires
freezing of the distortions due to only the $R^+_4$ mode. On the other
hand, because of the presence of an in-phase rotation along one axis,
the $a^+b^-b^-$ tilt system favored by the orthorhombic YNiO$_3$
requires freezing of the distortions due to both $R^+_4$ and $M^+_3$
modes. The value of the imaginary frequency of the $R^+_4$ mode
$\omega_{R^+_4} = 171i$ cm$^{-1}$ is noticeably larger than that of
the $M^+_3$ mode $\omega_{M^+_3} = 124i$ cm$^{-1}$ in LaNiO$_3$,
indicating that the distortions due to the $R^+_4$ mode might be more
favorable in this material. However, the instabilities of the $R^+_4$
and $M^+_3$ modes, with the respective frequencies $\omega_{R^+_4} =
299i$ and $\omega_{M^+_3} = 267i$ cm$^{-1}$, are much closer in
YNiO$_3$, suggesting that distortions due to both modes are likely to
occur in YNiO$_3$.  Thus the calculated phonon instabilities seemingly
provide the microscopic explanation for the different octahedral
rotations observed in LaNiO$_3$ and YNiO$_3$.

Although the calculated phonon instabilities are consistent with the
observed structural distortions in the two nickelates, these
instabilities could also lead to other structural distortions. For
example, phonon instabilities similar to that of LaNiO$_3$ occur in
SrTiO$_3$, but they cause out-of-phase rotations of the oxygen
octahedra about only one axis ($a^0a^0c^-$ in Glazer's notation) in
SrTiO$_3$ \cite{Lasota1997}. Howard and Stokes have shown that fifteen
different structures can arise out of the $R^+_4$ and $M^+_3$ phonon
instabilities \cite{Howard1998}. I generated all fifteen structures
for both LaNiO$_3$ and YNiO$_3$ and fully relaxed them using DFT
calculations within the PBE GGA. The energies of the relaxed
structures relative to that of the undistorted structure are given in
Table~\ref{tab:ene}.

\begin{table}
  \centering
  \caption{\label{tab:ene} The relative total energies of LaNiO$_3$
    and YNiO$_3$ with different Glazer tilts. The energies of the
    Glazer tilts that could not be stabilized are denoted by ``---''.}
  \vspace{0.1in}
  \begin{tabular}{l l d{1.3} d{1.3}}
    \toprule
                &             & \multicolumn{1}{c}{LaNiO$_3$}    & \multicolumn{1}{c}{YNiO$_3$}  \\
    tilt system & space group & \multicolumn{1}{c}{energy (meV/Ni)} & \multicolumn{1}{c}{energy (meV/Ni)}\\
    \midrule
    $a^0a^0a^0$  &  $Pm\overline{3}m$    &   0.0   &     0.0      \\
    $a^+a^+a^+$  &  $Im\overline{3}$     & -29.6   &  -461.3  \\
    $a^0b^+b^+$  &  $I4/mmm$             & -33.3   &  -489.7  \\
    $a^0a^0c^+$  &  $P4/mbm$             & -45.1   &  -491.0  \\
    $a^0a^0c^-$  &  $I4/mcm$             & -107.9  &  -613.4  \\
    $a^0b^-b^-$  &  $Imma$               & -114.6  &  -788.6  \\
    $a^-a^-a^-$  &  $R\overline{3}c$     & -115.9  &  -767.3  \\
    $a^+b^+c^+$  &  $Immm$               &  $---$  &   $---$  \\
    $a^+a^+c^-$  &  $P4_2/nmc$           &  $---$  &  -715.2  \\
    $a^0b^+c^-$  &  $Cmcm$               &  $---$  &  -731.3  \\
    $a^+b^-b^-$  &  $Pnma$               & -116.9  & -1010.1  \\
    $a^0b^-c^-$  &  $C2/m$               &  $---$  &   $---$  \\
    $a^-b^-b^-$  &  $C2/c$               &  $---$  &  -767.5  \\
    $a^+b^-c^-$  &  $P2_1/m$             &  $---$  &   $---$  \\
    $a^-b^-c^-$  &  $P\overline{1}$      &  $---$  &  -788.6  \\
    \bottomrule
  \end{tabular}
\end{table}

The calculations show that the gain in energy due to octahedral
rotations in LaNiO$_3$ is relatively small compared to that in
YNiO$_3$, which again confirms that LaNiO$_3$ is close to the cubic
phase. Not all octahedral tilt patterns could be stabilized, and these
structures are denoted by the symbol ``---'' in the table. The energy
of the $R\overline{3}c$ structure of LaNiO$_3$ with the tilt pattern
$a^-a^-a^-$ is $-$115.9 meV/Ni (i.e, per formula unit that consists of
five atoms) relative to that of the undistorted cubic
structure. Surprisingly, I find that the energy of the $Pnma$
structure with the tilt pattern $a^+b^-b^-$ to be even lower, albeit
by only 1.0 meV/Ni.  In addition, the structure with the tilt pattern
$a^0b^-b^-$ is only 1.3 meV/Ni higher in energy than the
$R\overline{3}c$ phase. 

The closeness in energy of several distinct structures of LaNiO$_3$
indicates that the structure of this material can dynamically
fluctuate at finite temperatures and suggests that the material might
be in the proximity of a structural quantum critical point.

All known diffraction experiments on powder and single crystal samples
of LaNiO$_3$ have found the structure to be rhombohedral with the
space group $R\overline{3}c$, but I find the orthorhombic structure
with the space group $Pnma$ to be lower in energy.  To check the
robustness of my calculations, I also did structrural relaxations
using other sets of pseudopotentials (GRVB LDA and ONCV PBE), and they
also give the lowest energy to the $Pnma$ phase. The only experiment
that is consistent with my finding is the pair density function
analysis of powder LaNiO$_3$ performed by Li \textit{et al.}\ who
found that the high-temperature phase of LaNiO$_3$ is best described
by an orthorhombic $Pnma$ structure at the nanoscale \cite{Li2016}. 

For YNiO$_3$, the orthorhombic $Pnma$ structure with the $a^+b^-b^-$
tilt pattern has the lowest energy. The energies of other tilt
patterns of YNiO$_3$ are much higher than the $Pnma$ structure, unlike
in the case of LaNiO$_3$. For example, the $Imma$ structure with the
$a^0b^-b^-$ tilt pattern, which is energetically closest to the $Pnma$
structure, is higher in energy by 221.5 meV/Ni. As noted above, the
imaginary frequencies of the $R^+_4$ and $M^+_3$ modes are closer in
YNiO$_3$ than in LaNiO$_3$. So the presence of competing structural
phases and possible proximity to a structural quantum critical point
is not caused by a near degeneracy of the $R^+_4$ and $M^+_3$ phonon
instabilities. The results shown here suggest that a larger difference
between the imaginary frequencies of the two phonon modes might lead
to such a competition, although a very large difference would probably
stabilize a structural distortion due to only one unstable mode.

\subsection{Spin-polarized structural relaxations}

Guo \textit{et al.}\ have recently reported an antiferromagnetic phase
transition at $T_N \sim 157$ K in their single crystal samples of
LaNiO$_3$ \cite{Guo2018}. The propagation wave vector observed for the
antiferromagnetic ordering in LaNiO$_3$ is
$(\ffrac14,\ffrac14,\ffrac14)_c$ in the cubic notation. In terms of
the reciprocal lattice vectors of the rhombohedral $R\overline{3}c$
and orthorhombic $Pnma$ unit cells, the propagating wave vectors are
$(\ffrac12,\ffrac12,\ffrac12)_r$ and $(\ffrac12,\ffrac12,0)_o$,
respectively.

Another recent experimental study on single crystal samples by Zhang
\textit{et al.}, however, did not find any such transition
\cite{Zhang2017}. A broad feature in the susceptibility measurements
is seen in Zhang \textit{et al.}'s samples and a small anomaly
had also been observed previously \cite{Li2016}, which suggests that
long range magnetic ordering might only occur in highly pure samples.


To understand the nature of magnetic instabilities, if there are any,
in LaNiO$_3$ and possible competition between different magnetic
interactions, I extensively studied the stability of diverse magnetic
ordering phases in several supercells of $R\overline{3}c$ and $Pnma$
structures using spin-polarized DFT calculations within the PBE GGA.

\subsubsection{Spin-polarized structural relaxations in
  $R\overline{3}c$ LaNiO$_3$}

In $R\overline{3}c$ LaNiO$_3$, I found a weak ferromagnetic
instability with an ordered moment of 0.2 $\mu_B$/Ni and an energy
gain of 0.3 meV/Ni relative to the paramagnetic state. I was not able
to stabilize the $A$-, $C$-, or $G$-type orderings in a 40-atom
supercell of the $R\overline{3}c$ structure. All these orderings
showed negligible moments and no discernible gain in energy.

The experimentally observed antiferromagnetic order corresponds to an
80-atom $2\times2\times2$ supercell of the $R\overline{3}c$
structure. Since the $R\overline{3}c$ unit cell has two formula units,
there are sixteen Ni atoms in this supercell. The propagation vector
uniquely determines the ordering pattern of eight of these Ni
atoms. The ordering pattern in the sublattice formed by the remaining
eight atoms is also uniquely determined by the propagation vector, and
each Ni atom in one sublattice has three spin-up and three spin-down
Ni sites of the other sublattice as its nearest neighbors.  So there
is only one collinear antiferromagnetic ordering pattern consistent
with the reported ordering wave vector in this
$R\overline{3}c$-derived supercell. However, the two sublattices have
the freedom to have different values for the on-site magnetic
moments. In case such an antiferromagnetic state with different
on-site moment occurs, it would correspond to a rock-salt ordering of
the breathing distortions in the rhombohedral phase similar to the one
observed for the orthorhombic rare-earth nickelates.  However, I was
not able to stabilize any magnetic orderings with noticeable magnetic
moments and energy gains in this 80-atom unit cell.

I also tried to stabilize antiferromagnetic orderings in the 20-atom
$1\times1\times2$ supercell corresponding to the propagation wave
vector $(0,0,\ffrac12)_r$ of the $R\overline{3}c$ unit cell. This wave
vector is equivalent to $(\ffrac14,\ffrac14,\ffrac34)_c$ in the
pseudocubic notation, and it is also compatible with the neutron
scattering experiments of Guo \textit{et al.} \cite{Guo2018}. In this
supercell, the ordering wave vector uniquely determines the collinear
antiferromagnetic ordering pattern up to any disproportionation. The
wave vector partitions the four Ni atoms in this supercell into two
sublattices with two Ni each, and each Ni site in one sublattice has
three spin-up and three spin-down Ni sites from the other sublattice
as its nearest neighbors. The spins are ordered ferromagnetically in
the $ab$ plane and antiferromagnetically along the $(0,0,1)_r$
direction.  I was able to obtain such an antiferromagnetic solution
with an energy gain of 0.7 meV/Ni relative to the nonmagnetic
state. The two Ni sublattices have slightly different magnitudes of
0.27 and 0.26 $\mu_B$ for the magnetic moments, so this phase already
shows a tendency towards disproportionation.  The disproportionated
state shows an even larger energy gain of 2.1 meV/Ni relative to the
nonmagnetic phase and has Ni sites with magnetic moments of 0.6
$\mu_B$ and zero inside the larger and smaller octahedra,
respectively. The disproportionated octahedra are arranged in a
rock-salt-type arrangement, consistent with Mazin \textit{et al.}'s
picture of charge ordering \cite{Mazin2007}.

The presence of the breathing distortions in the
$R\overline{3}c$-derived supercell is, however, inconsistent with Lee
\textit{et al.}'s symmetry analysis within a Landau theory, which
suggested that no charge ordering occurs in the antiferromagnetic
ordering of the rhombohedral structure \cite{Lee2011a}.

\begin{figure}
  \centering
  \includegraphics[width=0.7\textwidth]{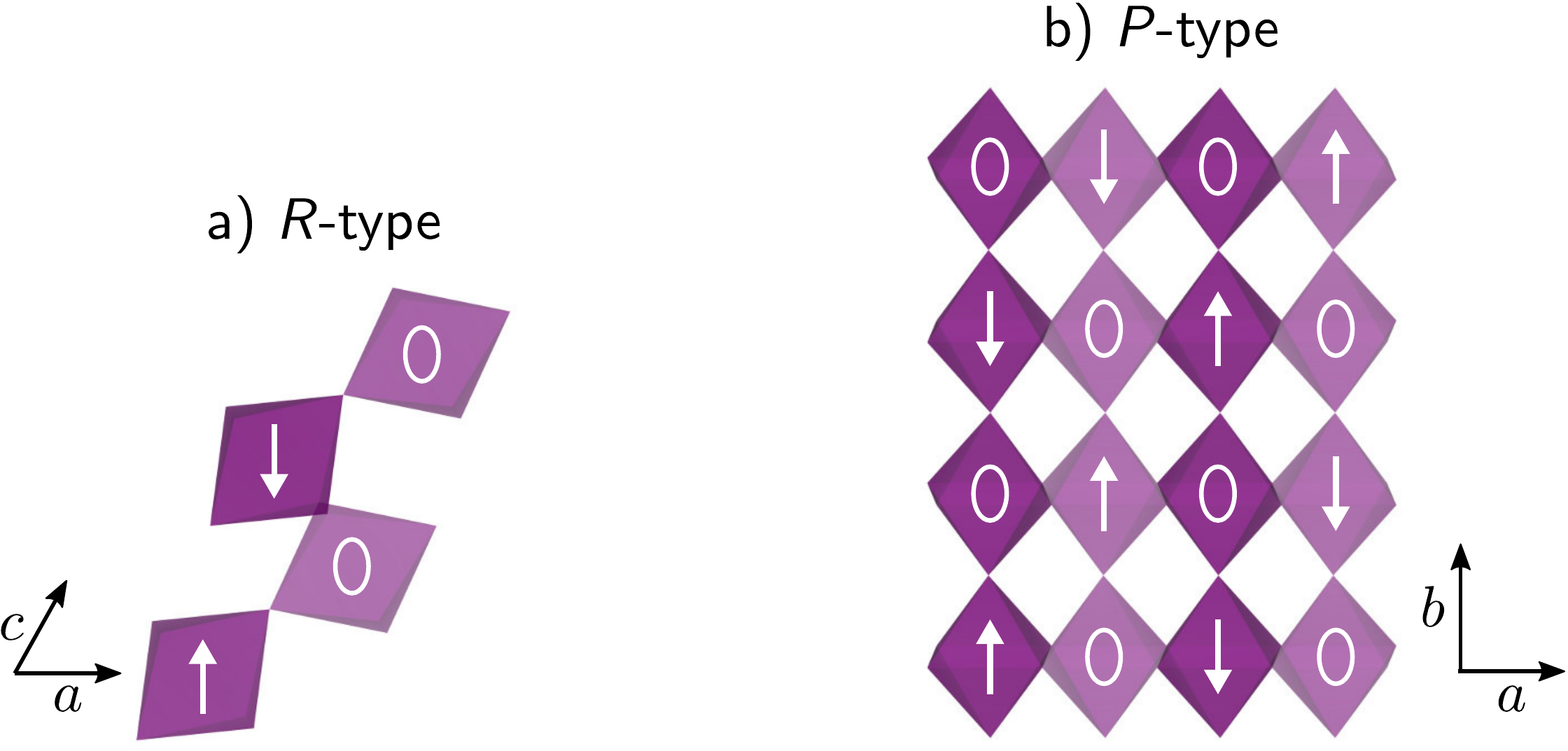}
  \caption{ The a) $R$- and b) $P$-type phases derived from the
    $R\overline{3}c$ and $Pnma$ structures with the propagation wave
    vectors $(0,0,\ffrac12)_r$ and $(\ffrac12,\ffrac12,0)_o$,
    respectively. Only the smallest units of repetition are shown. The
    octahedra in different layers in the out-of-plane direction are
    shaded differently. The octahedra with the symbol ``0'' have
    smaller volumes, and the Ni sites inside them have no magnetic
    moments.}
  \label{fig:afm} 
\end{figure}

Although the difference in the magnitude of the magnetic moments is
large in the disproportionated phase, the breathing distortions are
relatively small. The larger octahedra have a volume of 10.1 \AA$^3$,
while the smaller ones have a volume of 9.8 \AA$^3$. The Ni-O bond
lengths in the two sets of octahedra are 1.95 and 1.96 \AA,
respectively, and this small difference of 0.01 \AA\ might be the
reason for the difficulty in observing this distortion in the
experiments. This disproportionated antiferromagnetic state is labeled
as $R$-type in this work and is illustrated in Fig.~\ref{fig:afm}(a).

\subsubsection{Spin-polarized structural relaxations in
  $Pnma$ LaNiO$_3$}

Since the orthorhombic $Pnma$ phase of LaNiO$_3$ has a slightly lower
energy than the $R\overline{3}c$ phase in my calculations, I also
explored if antiferromagnetism occurs in the orthorhombic phase. I was
able to stabilize a ferromagnetic state with an ordered moment of 0.2
$\mu_B$/Ni and an energy gain of 0.6 meV/Ni relative to the
nonmagnetic phase. But I was not able to stabilize the $A$-, $C$-, or
$G$-type ordering arrangements.

I constructed an 80-atom $2\times2\times1$ supercell of the $Pnma$
structure that corresponds to the experimentally observed propagation
wave vector of the antiferromagnetic order. The $Pnma$ unit cell has
four formula units, so this supercell has sixteen Ni atoms. The
constraint due to the propagation wave vector partitions the Ni
lattice into four sublattices with four Ni atoms each. One can
enumerate eight arrangements of collinear spin orderings within this
constraint, but only two of them are symmetrically inequivalent.
These are the so-called $S$- and $T$-type orderings
\cite{Giovannetti2009}.  In both these orderings, the spins are
ordered $\uparrow\uparrow\downarrow\downarrow$ in the $ac$ plane and
are flipped to $\downarrow\downarrow\uparrow\uparrow$ in the
next-nearest plane along the $b$ direction.  They are sandwiched by
nearest-neighbor planes with the same types of spin arrangements in
the $S$-type ordering, while in the $T$-type ordering, they are
sandwiched by layers with the spin orderings
$\uparrow\downarrow\downarrow\uparrow$ and
$\downarrow\uparrow\uparrow\downarrow$. Like the spin orderings in the
rhombohedral supercells that are compatible with the experimentally
observed propagating wave vector, each Ni atom in these spin orderings
have three spin-up and spin-down Ni atoms as their nearest
neighbors. In addition, the four Ni sublattices in the supercell have
the freedom to disproportionate and have different on-site moments.

I was able to stabilize both the $S$- and $T$-type antiferromagnetic
orderings in the $2\times2\times1$ supercell of the orthorhombic
LaNiO$_3$. The energy gain of $\sim$0.4 meV/Ni relative to the
nonmagnetic phase due to these orderings is small, like in the
nondisproportionated antiferromagnetic state corresponding to the wave
vector $(0,0,\ffrac12)_r$ of the $R\overline{3}c$ unit cell. The
on-site Ni moments are $\sim$0.2 $\mu_B$ in the $S$- and $T$-type
antiferromagnetic phases, but the magnitudes of the moments vary by
$\sim$5\% in different Ni sublattices. I found that both these
orderings show a strong propensity to disproportionate in a rock-salt
pattern of alternating large and small NiO$_6$ octahedra. The Ni sites
inside the large octahedra have a moment of 0.6 $\mu_B$, while the
ones inside the small octahedra have no magnetic moment. The
disproportionated phases of the $S$- and $T$-type orderings are
symmetrically identical, and this phase has an energy gain of 2.0
meV/Ni relative to the nonmagnetic phase. This phase is labeled as
$P$-type and is shown in Fig.~\ref{fig:afm}(b). The larger and smaller
octahedra in this structure have volumes of 10.2 and 9.9 \AA$^3$,
respectively. The Ni-O distances in the corresponding octahedra are
1.97 and 1.95 \AA. These values are similar to the one obtained for
the $R\overline{3}c$-derived $R$-type phase. In fact, the $P$- and
$R$-type phases are only distinguished by their underlying crystal
structures. The magnetic ordering is same in these phases, with the Ni
sites with zero moments having three spin-up and three spin-down
nearest neighbors.

\subsection{Lindhard susceptibility and antiferromagnetic ordering} 

All three nondisproportionated antiferromagnetic solutions that I
obtained for LaNiO$_3$ exhibit a propensity for the octahedral
breathing distortions.  This supports Mazin \textit{et al.}'s theory
that the nickelates occur in a crossover between the localized and
itinerant regimes where the nearest-neighbor Ni sites have a tendency
to disproportionate \cite{Mazin2007}. However, the 80-atom supercell
of the $R\overline{3}c$ structure corresponding to the propagation
vector $(\ffrac12,\ffrac12,\ffrac12)_r$ also allows simultaneous
existence of both antiferromagnetism and rock-salt-type
disproportionation. But I could not stabilize this phase. To determine
if there is any connection between the Fermi surface instabilities and
the magnetic orderings, I calculated the Lindhard susceptibility
\[
\chi_0(q,\omega) = \sum_{k,m,n} |M_{k,k+q}^{m,n}|^2
\frac{f(\epsilon_k^m) - f(\epsilon_{k+q}^n)}{\epsilon_k^m -
  \epsilon_{k+q}^n - \omega - \imath \delta}
\]
at $\omega \to 0$ and $\delta \to 0$, where $\epsilon_k^m$ is the
energy of a band $m$ at the wave vector $k$ and $f$ is the Fermi
distribution function. $M$ is the matrix element, which is set to
unity for the constant matrix element approximation employed here. The
negligence of the matrix element changes the relative intensities of
the peaks in the susceptibility \cite{Heil2014}. However, major
features remain the same and qualitative understanding can still be
gleaned off from such an approximation. I note that a previous
discussion of the magnetic susceptibility in the nickelates has also
made this approximation \cite{Lee2011b}.

\begin{figure}
  \centering
  \includegraphics[width=\textwidth]{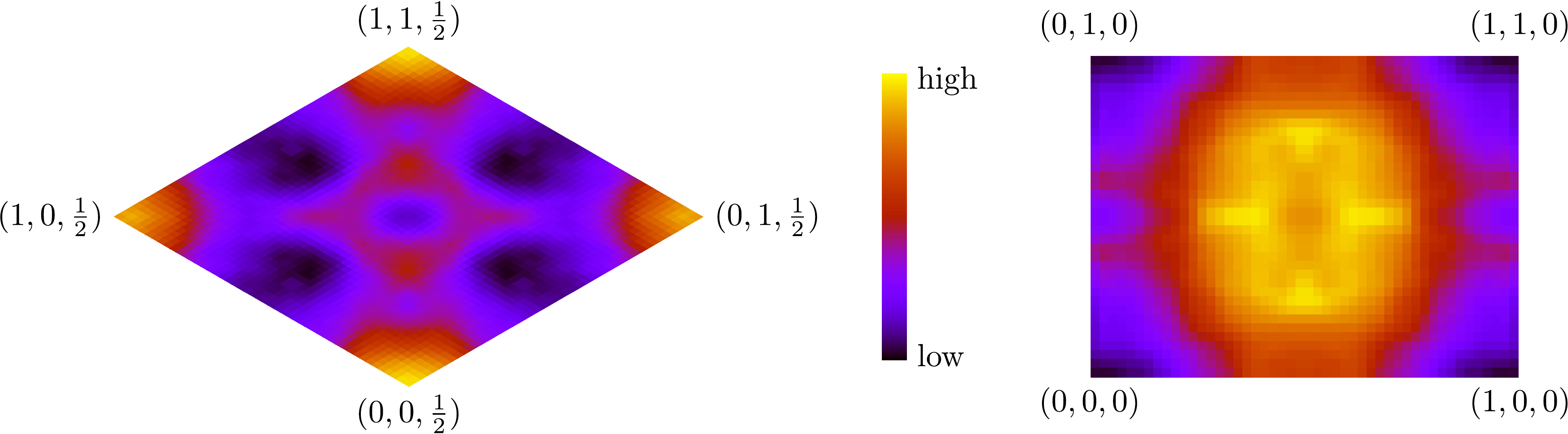}
  \caption{Calculated Lindhard susceptibility of LaNiO$_3$ in the
    $R\overline{3}c$ (left) and $Pnma$ (right) structures. }
  \label{fig:lind}
\end{figure}

The calculated Lindhard susceptibility of the $R\overline{3}c$ phase
for the $(q_a,q_b,\ffrac12)_r$ plane is shown in
Fig.~\ref{fig:lind}(a). The susceptibility shows a peak at
$(0,0,\ffrac12)_r$, the propagation wave vector for which I was able
to stabilize an antiferromagnetic solution. The calculated
susceptibility at $(\ffrac12,\ffrac12,\ffrac12)_r$ is low and occurs
at a local minima, and this seems to explain the lack of
antiferromagnetic instability at this wave vector in the
calculations. This result is consistent with Lee \textit{et al.}'s
suggestion that Fermi surface nesting plays an important role in the
antiferromagnetic instability of the rare-earth nickelates
\cite{Lee2011a,Lee2011b}. [But note the discussion in the following
  paragraph.] However, it is the nesting instability of the
rhombohedral, not the cubic, phase that is important because a peak at
the wave vector $(\ffrac14,\ffrac14,\ffrac14)_c =
(\ffrac12,\ffrac12,\ffrac12)_r$ does occur in the cubic phase
\cite{Lee2011a,Lee2011b}.

The calculated Lindhard susceptibility of the orthorhombic $Pnma$
LaNiO$_3$, which is shown in Fig.~\ref{fig:lind}(b), does not exhibit
a sharp peak at $(\ffrac12,\ffrac12,0)_o$ that corresponds to the
experimentally observed ordering wave vector. Instead of sharp peaks,
the susceptibility of the $Pnma$ phase shows a plateau-like
enhancement of the susceptibility in the region $(\ffrac14 < q_a
<\ffrac34, 0 < q_b < 1,0)_o$. The wave vector
$(\ffrac12,\ffrac12,0)_o$ actually occurs at a local
minimum. Therefore, although a high Lindhard susceptibility appears
necessary for the disproportionated antiferromagnetic instabilities,
a sharp peak corresponding to a well-defined nesting does not seem to
be crucial. Importantly, because there are four Ni sublattices, the
antiferromagnetic orderings in the 80-atom supercell corresponding to
the wave vector $(\ffrac12,\ffrac12,0)_o$ can host either the
nearest-neighbor rock-salt-type disproportionation or the layered
disproportionation where the neighboring Ni planes are alternatingly
disproportionated. However, I was not able to stabilize the latter
arrangement of disproportionation, which shows that the
nearest-neighbor rock-salt ordering of the breathing distortions play
an essential role in the phase transition of LaNiO$_3$.

The sizable energy gain due to rock-salt-type disproportionations, and
the lack of such a gain in other arrangements of the breathing
distortions, indicates that the structural instability that leads to
the breathing distortions is not just related to the moment
formation. In addition to the moment formation, my calculations
suggest that the breathing instability arises due to coupling with the
antiferromagnetic fluctuations that involve three spin-up and
spin-down nearest neighbors. This is also supported by the fact that I
was not able to stabilize breathing distortions in the
non-spin-polarized calculations.


\begin{figure}
  \includegraphics[width=0.5\textwidth]{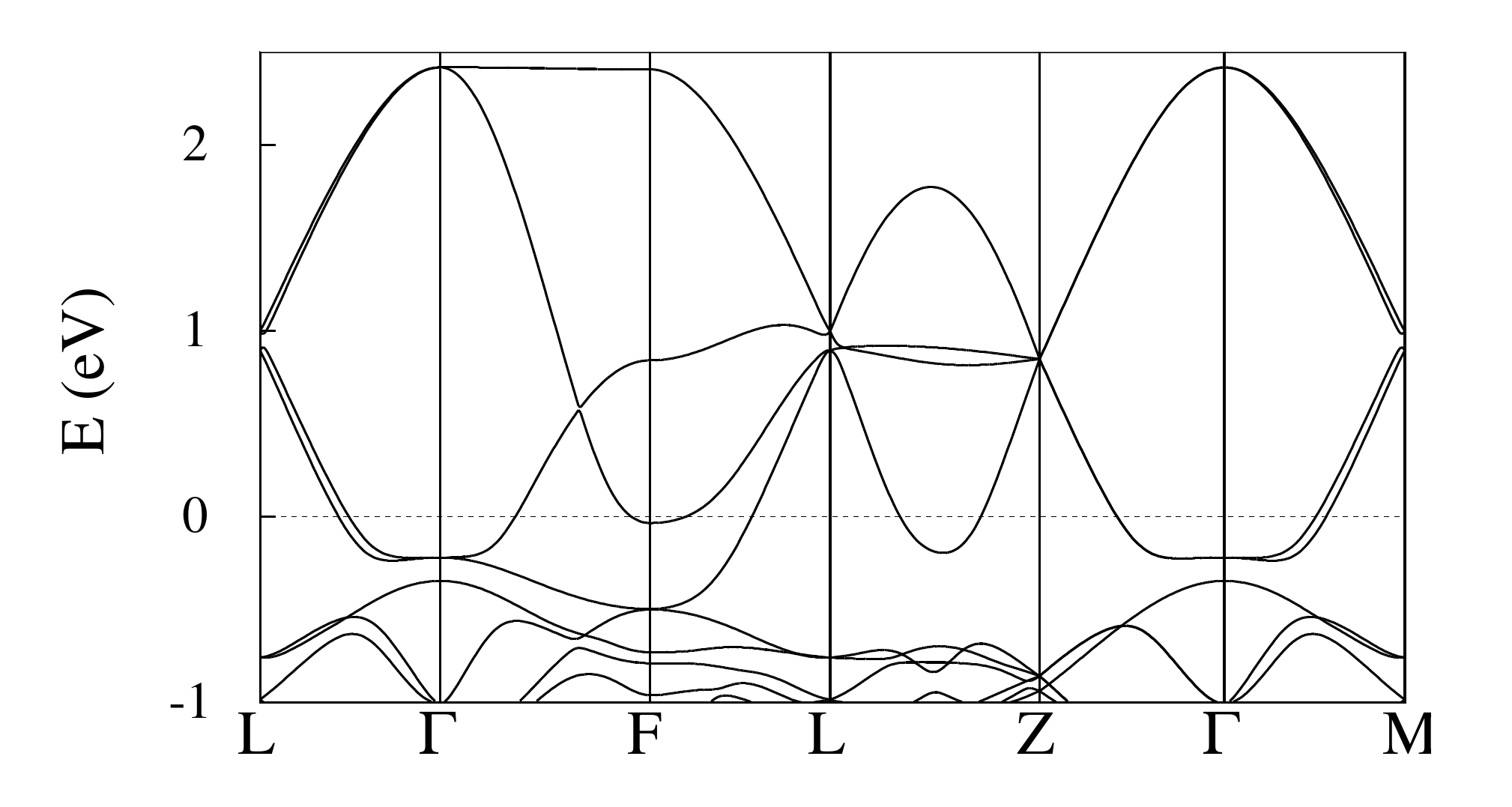}
  \includegraphics[width=0.5\textwidth]{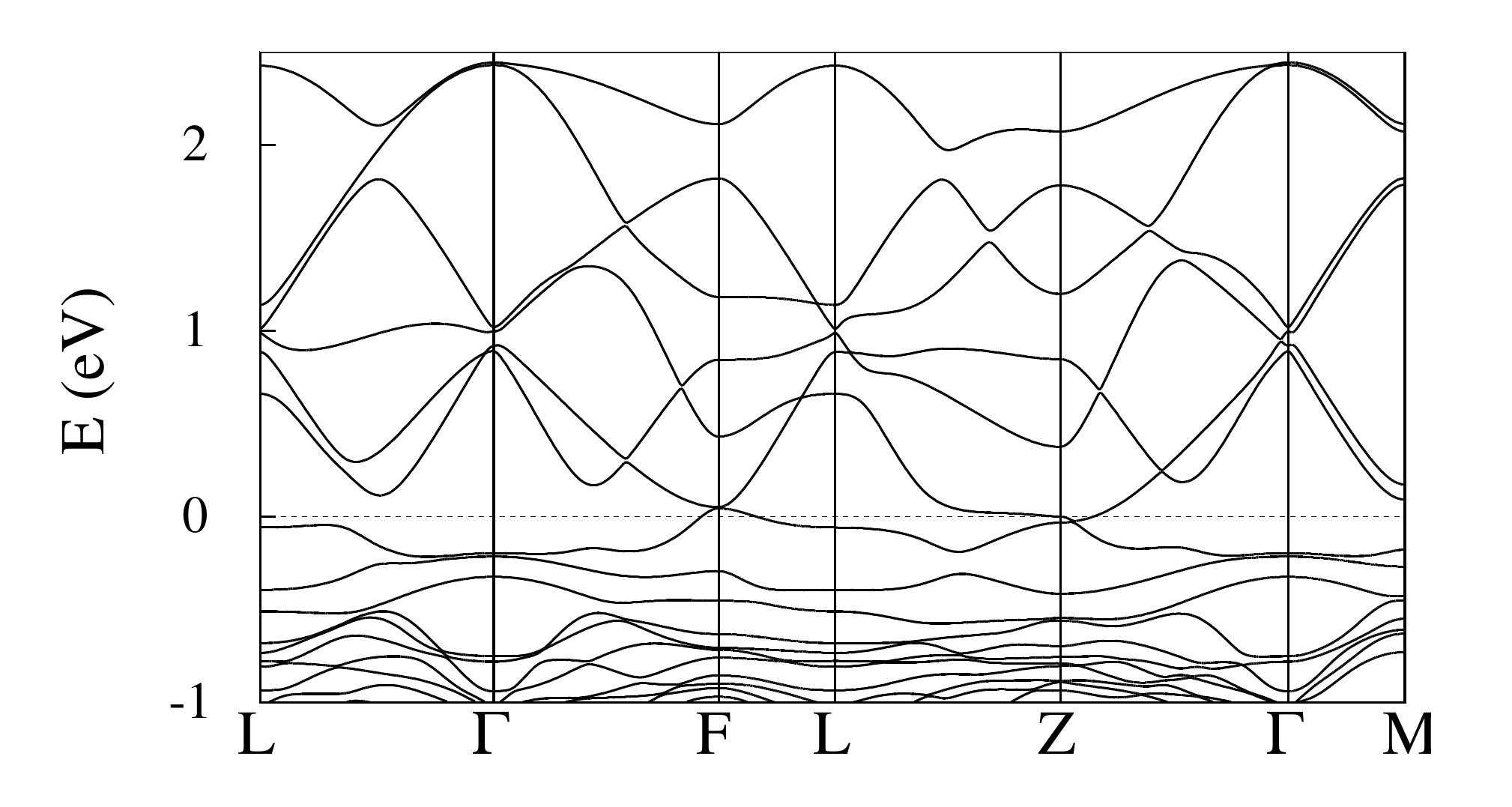}
  \caption{Calculated band structures of the nonmagnetic
    $R\overline{3}c$ LaNiO$_3$ (left) and the disproportionated
    antiferromagnetic $R$-type LaNiO$_3$ corresponding to the
    propagation wave vector $(0,0,\ffrac12)_r$. The band structures
    are plotted along the path $L\ (\ffrac12,0,0)_r \rightarrow
    \Gamma\ (0,0,0) \rightarrow F\ (\ffrac12,\ffrac12,0)_r \rightarrow
    L\ (\ffrac12,0,0)_r \rightarrow Z\ (\ffrac12,\ffrac12,\ffrac12)_r
    \rightarrow \Gamma\ (0,0,0) \rightarrow M\ (0,0,\ffrac12)_r$. The
    coordinates are given in terms of the reciprocal lattice vectors
    of the primitive cell.}
  \label{fig:rhomb-bs}
\end{figure}

\begin{figure}
  \includegraphics[width=0.5\textwidth]{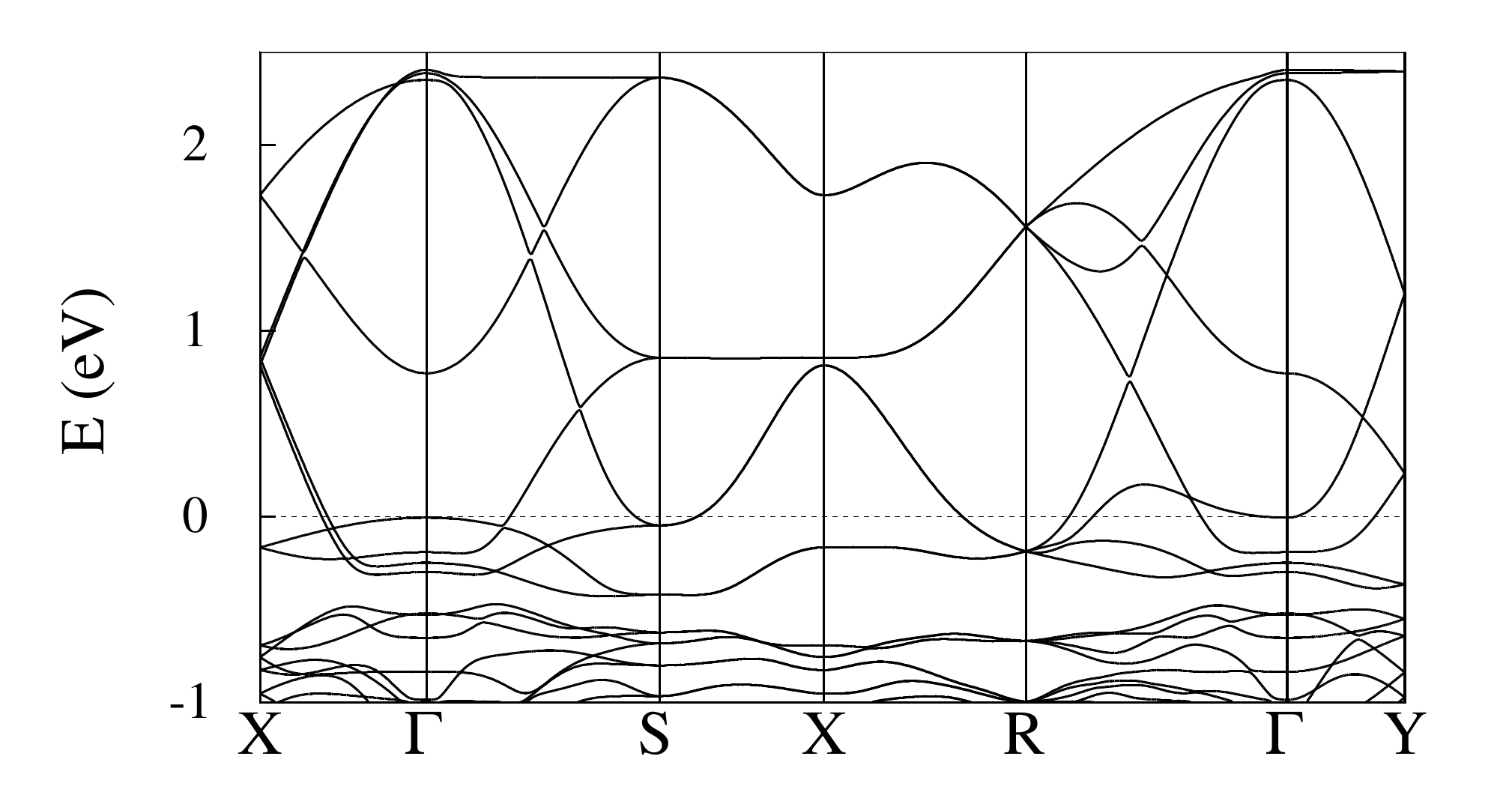}
  \includegraphics[width=0.5\textwidth]{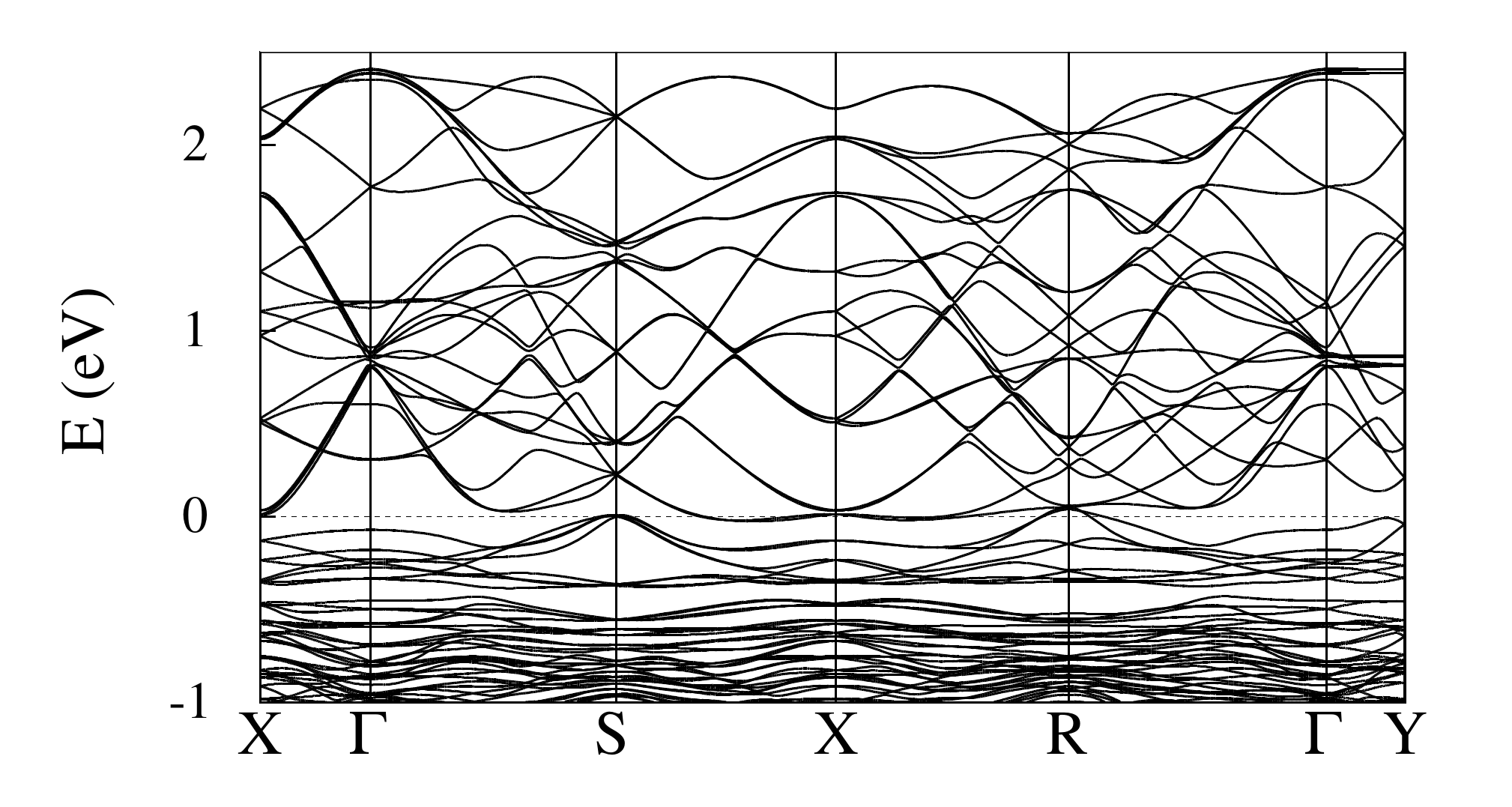}
  \caption{Calculated band structures of the nonmagnetic $Pnma$
    LaNiO$_3$ (left) and the disproportionated antiferromagnetic
    $P$-type LaNiO$_3$ corresponding to the propagation wave vector
    $(\ffrac12,\ffrac12,0)_o$. The band structures are plotted along
    the path $X\ (\ffrac12,0,0)_o \rightarrow \Gamma\ (0,0,0)
    \rightarrow S\ (\ffrac12,0,\ffrac12)_o \rightarrow
    X\ (\ffrac12,0,0)_o \rightarrow R\ (\ffrac12,\ffrac12,\ffrac12)_o
    \rightarrow \Gamma\ (0,0,0) \rightarrow Y\ (0,\ffrac12,0)_o$. The
    coordinates are given in terms of the reciprocal lattice vectors
    of the primitive cell.}
  \label{fig:ortho-bs}
\end{figure}

\begin{figure}
  \includegraphics[width=0.5\textwidth]{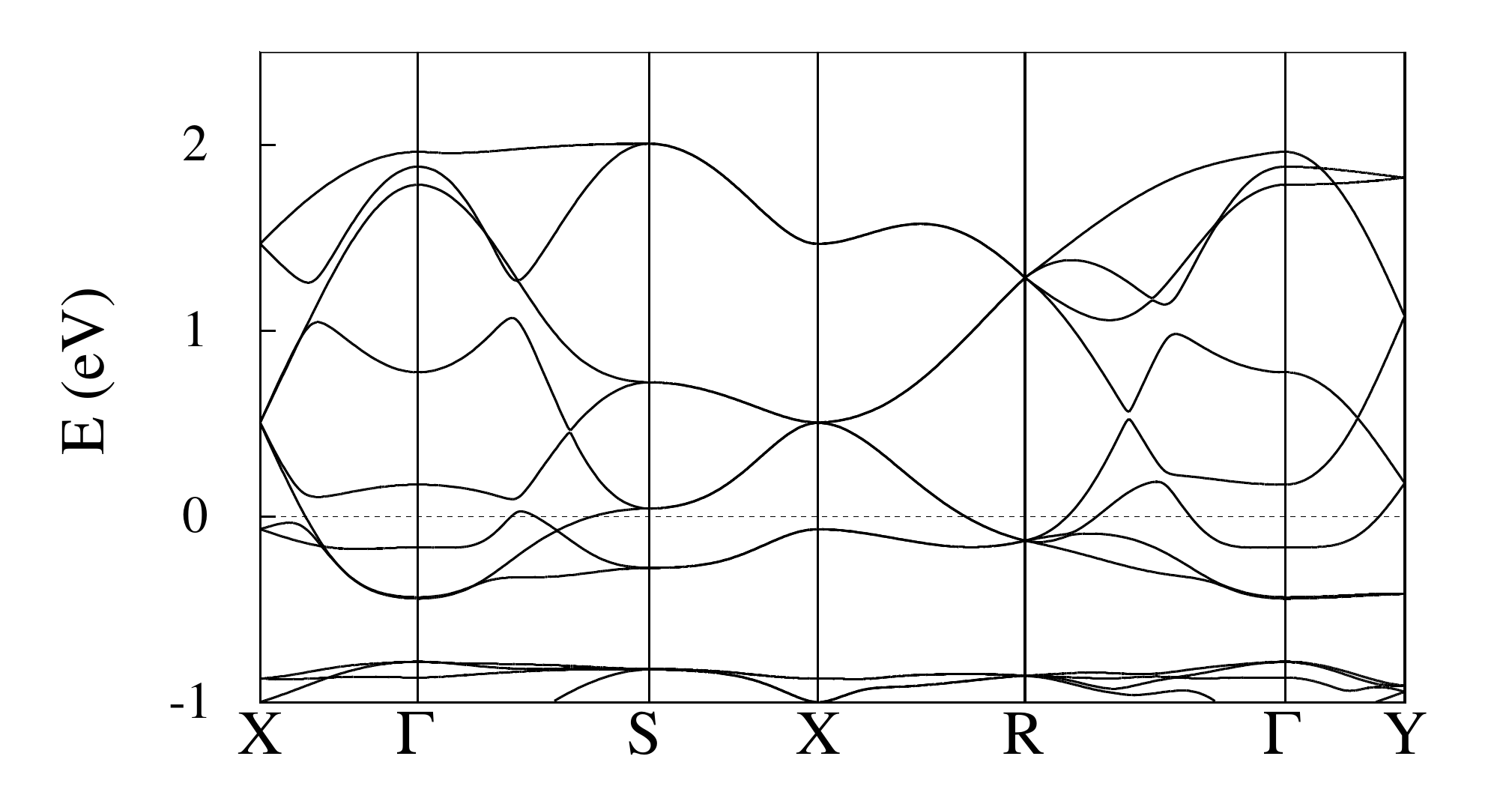}
  \includegraphics[width=0.5\textwidth]{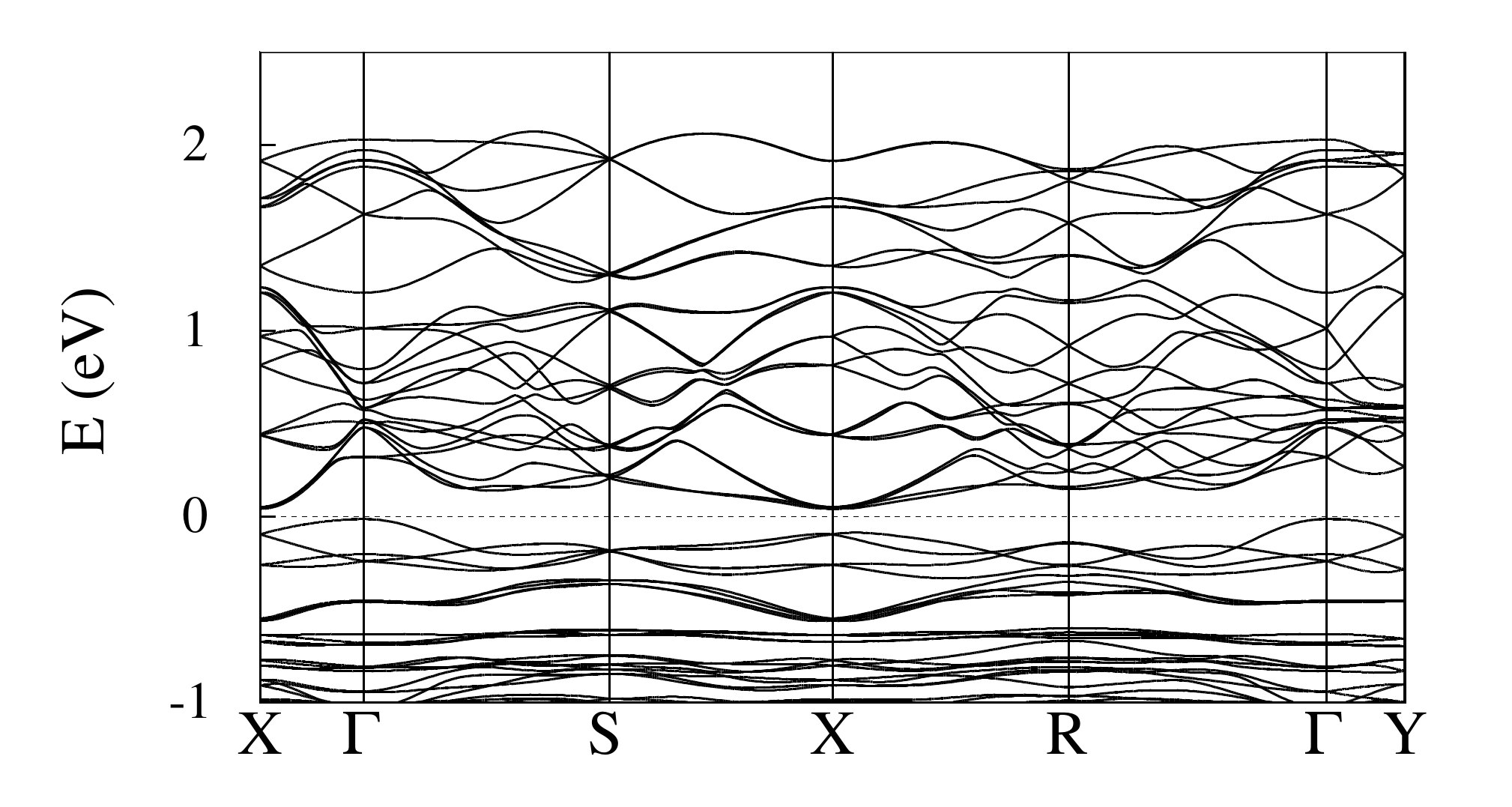}
  \caption{Calculated band structures of the nonmagnetic $Pnma$
    YNiO$_3$ (left) and the disproportionated antiferromagnetic
    $P$-type YNiO$_3$ corresponding to the propagation wave vector
    $(\ffrac12,\ffrac12,0)_o$. The path is as in
    Fig.~\ref{fig:ortho-bs}.}
  \label{fig:yno-bs}
\end{figure}

\begin{figure}
  \includegraphics[width=0.5\textwidth]{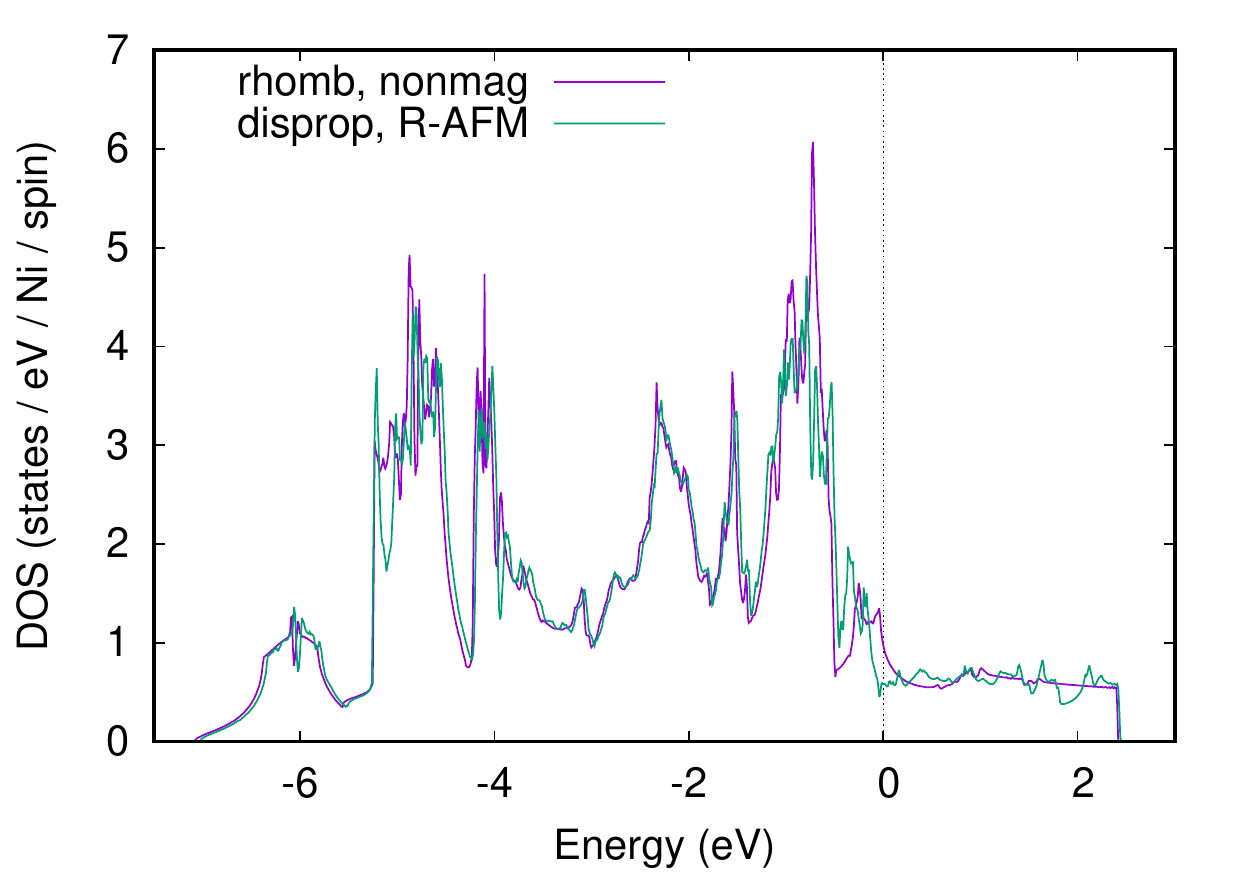}
  \includegraphics[width=0.5\textwidth]{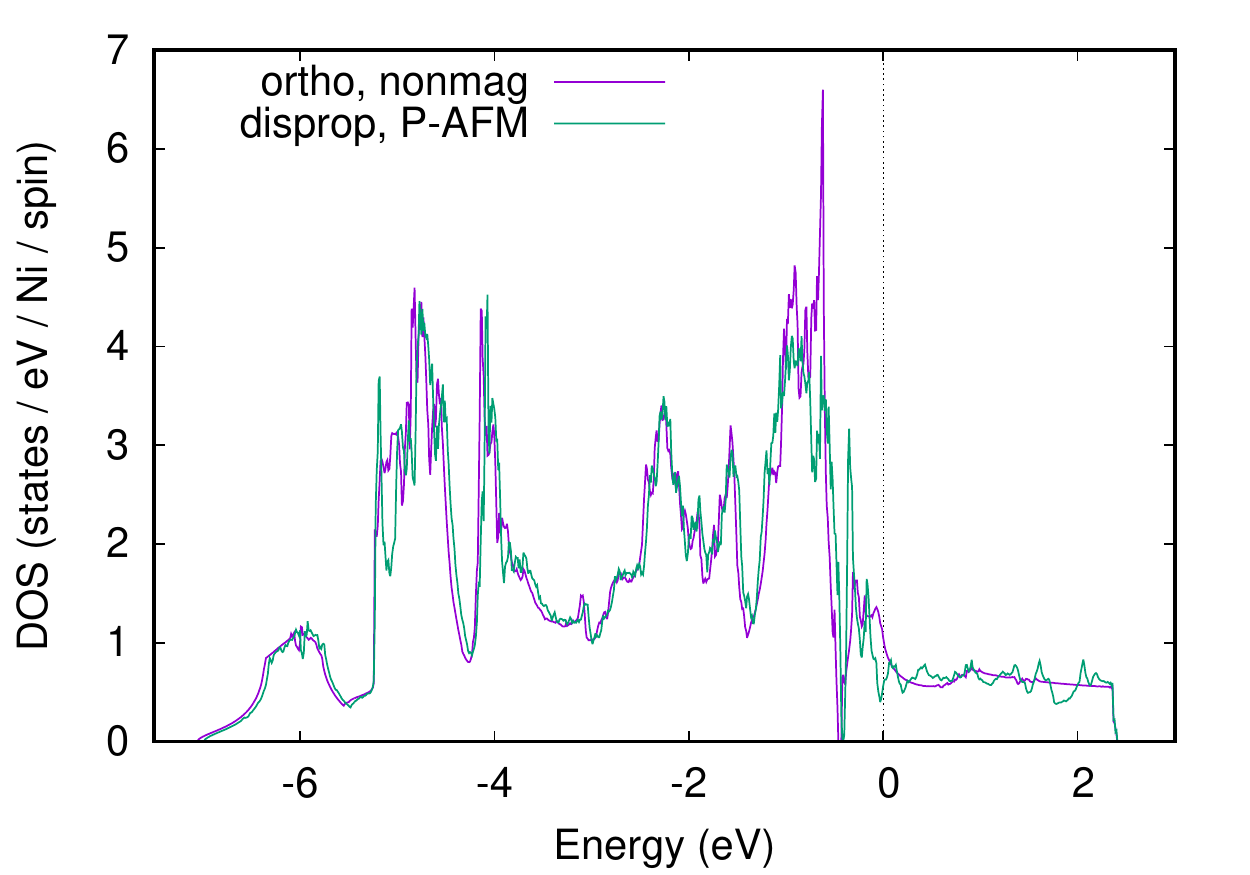}
  \caption{Left: Calculated electronic DOS of LaNiO$_3$ in the
    rhombohedral non-magnetic and disproportionated antiferromagnetic
    $R$-type structures. Right: DOS of LaNiO$_3$ in the orthorhombic
    non-magnetic and disproportionated antiferromagnetic $P$-type
    structures.}
  \label{fig:dos}
\end{figure}

\subsection{Electronic structure of the disproportionated
  antiferromagnetic phases}

The calculated band structures of LaNiO$_3$ in the nonmagnetic
$R\overline{3}c$ and $Pnma$ structures without the disproportionation
and in the corresponding low-symmetry $R$- and $P$-type
antiferromagnetic phases with the breathing distortions are shown in
Figs.~\ref{fig:rhomb-bs} and \ref{fig:ortho-bs}, respectively. For
comparison, the band structures of YNiO$_3$ in the nonmagnetic and
$P$-type phases are also shown in Fig.~\ref{fig:yno-bs}.

From the band structures of the nonmagnetic phases, one can readily
identify the antibonding $e_g$-derived bands between $-$0.5 and 2.5 eV
relative to the Fermi level. There are two spin-degenerate $e_g$ bands
per Ni in each structures. For example, the $R\overline{3}c$ structure
with two Ni per primitive cell has four bands in this manifold
[Fig.~\ref{fig:rhomb-bs}(left)], while the $Pnma$ structure with four
Ni per primitive cell has eight bands
[Fig.~\ref{fig:ortho-bs}(left)]. This $e_g$-derived manifold is
completely separated in the $Pnma$ structure, but it touches a lower
manifold in the $R\overline{3}c$ structure. [Such a crossing is
  allowed in the $R\overline{3}c$ structure because the bands
  belonging to the lower manifold also have the $e_g$ representation
  in this space group.] The Fermi level lies near the bottom of this
manifold and corresponds to a quarter filling.  The Fermi surfaces in
these phases are large and consist of multiple sheets (see
Figs.~\ref{fig:rhomb-fs} and \ref{fig:ortho-fs} in the appendix).
The electronic density of states (DOS) at the Fermi level is 1.0
states eV$^{-1}$ per Ni per spin in both the $R\overline{3}c$ and
$Pnma$ structures. This corresponds to a Sommerfeld coefficient of
$\gamma =$ 4.7 mJ mol$^{-1}$ K$^{-2}$, which is about 3.6 times
smaller than the experimentally determined value of 17 mJ mol$^{-1}$
K$^{-2}$ \cite{Guo2018,Zhou2014}.

As can be seen in the right panels of Figs.~\ref{fig:rhomb-bs} and
\ref{fig:ortho-bs}, the antiferromagnetic ordering and breathing
distortions in LaNiO$_3$ have a drastic effect only on the lower half
of the $e_g$-derived manifold. This is also apparent from the DOS plot
shown in Fig.~\ref{fig:dos} that reveals a large decrease of the DOS
value at the Fermi energy due to a downward movement of a peak in the
low-symmetry phases.

In the nonmagnetic phases, the lower and upper halves of the
$e_g$-derived manifold touch at isolated points in the Brillouin
zone. If one considers the two halves of the $e_g$-derived manifold as
either separate or weakly coupled to each other, the lower half with
one $e_g$ band per Ni is nominally half filled, and the
antiferromagnetic ordering splits apart this half-filled manifold.
For example, Fig.~\ref{fig:rhomb-bs}(right) shows the band structure
of the disproportionated antiferromagnetic $R$-type ordered phase of
LaNiO$_3$, which occurs in a $1\times1\times2$ supercell of the
nonmagnetic phase corresponding to the propagation wave vector
$(0,0,\ffrac12)_r$. The number of bands are now doubled to eight in
the $e_g$-derived manifold because there are four Ni atoms in this
supercell. Between $\sim$1 and 2.5 eV, i.e.\ in the upper half of the
$e_g$-derived manifold, the transition to the antiferromagnetic and
disproportionated phase does not induce any dramatic gaps. The four
bands in this upper half develop small gaps to remove the degeneracies
that arise from band foldings, but these four bands largely exhibit
the imprint of the two highest bands of the $R\overline{3}c$
structure. However, a large rupture appears in the lower half of the
$e_g$-derived manifold. Two bands are shifted above the Fermi level
and two bands below it, and the electronic structure near the Fermi
level looks nothing like that of the nonmagnetic phase without the
disproportionation.

\begin{figure}
  \centering
  \includegraphics[width=0.4\textwidth]{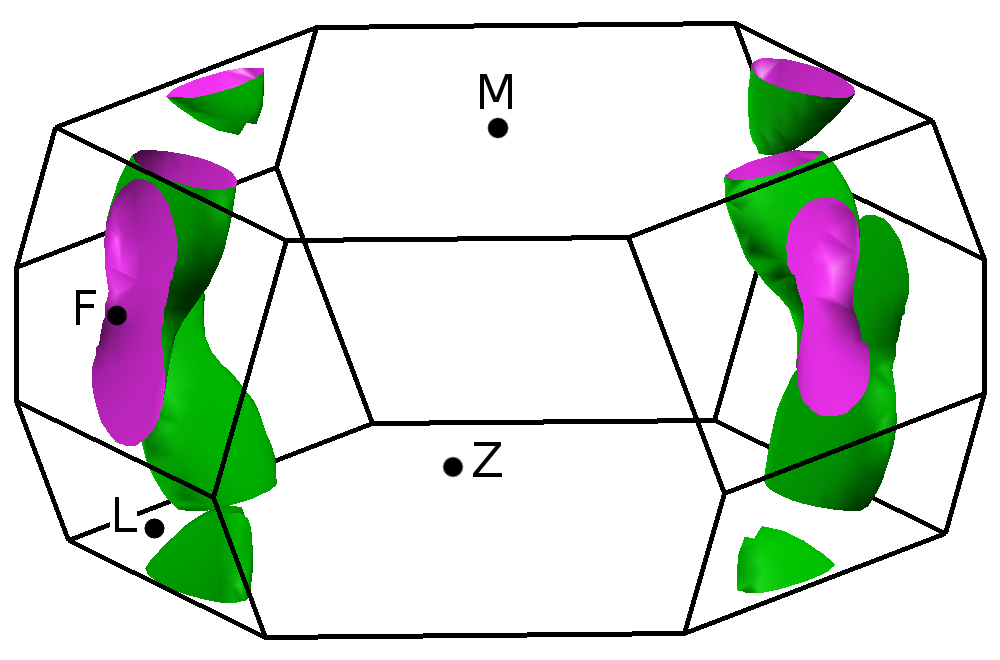}
  \qquad
  \includegraphics[width=0.4\textwidth]{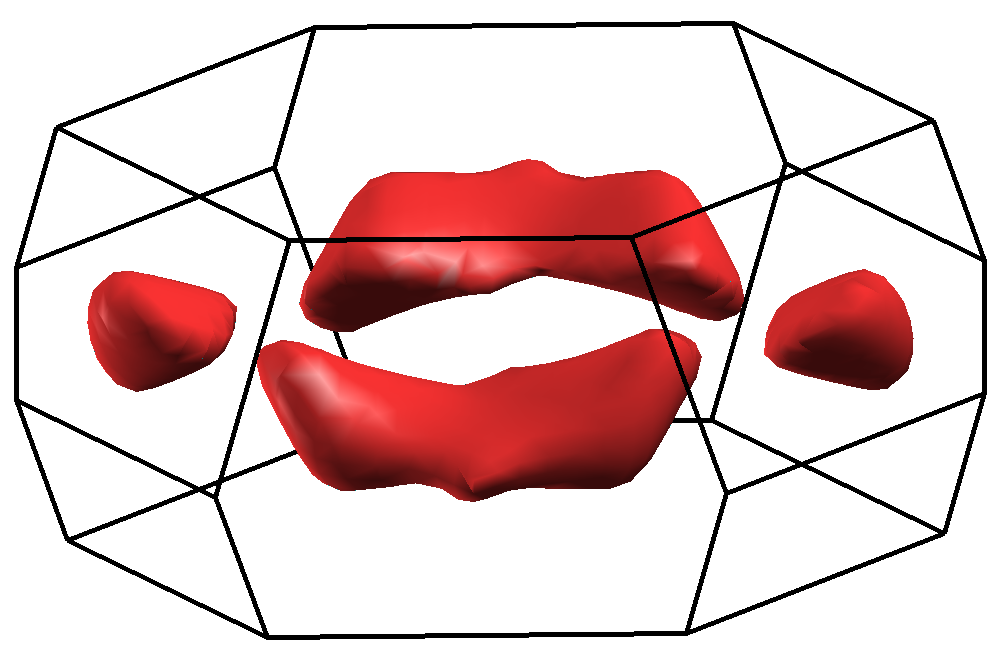}
  \caption{Left: Calculated Fermi sheets of LaNiO$_3$ in the
    disproportionated antiferromagnetic $R$-type phase.}
  \label{fig:rhomb-br-fs}
\end{figure}

\begin{figure}[!t]
  \centering
  \includegraphics[width=0.4\textwidth]{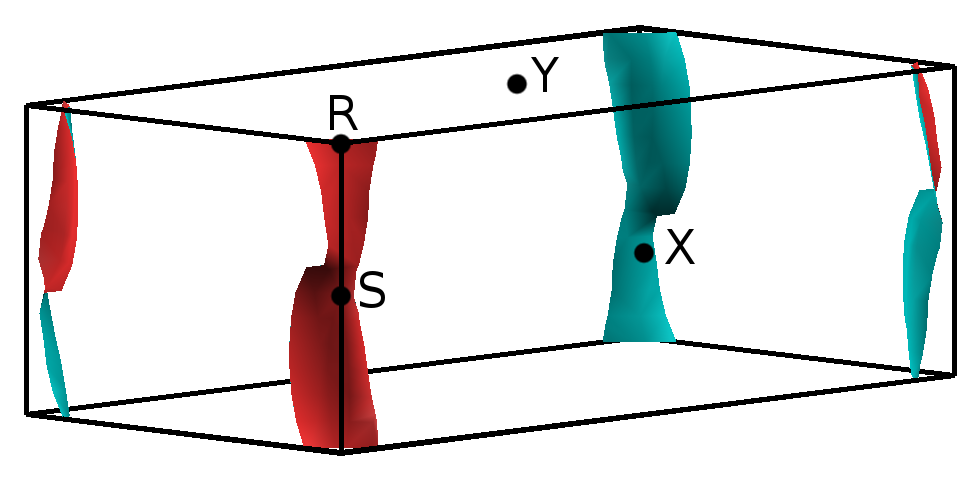}
  \qquad
  \includegraphics[width=0.4\textwidth]{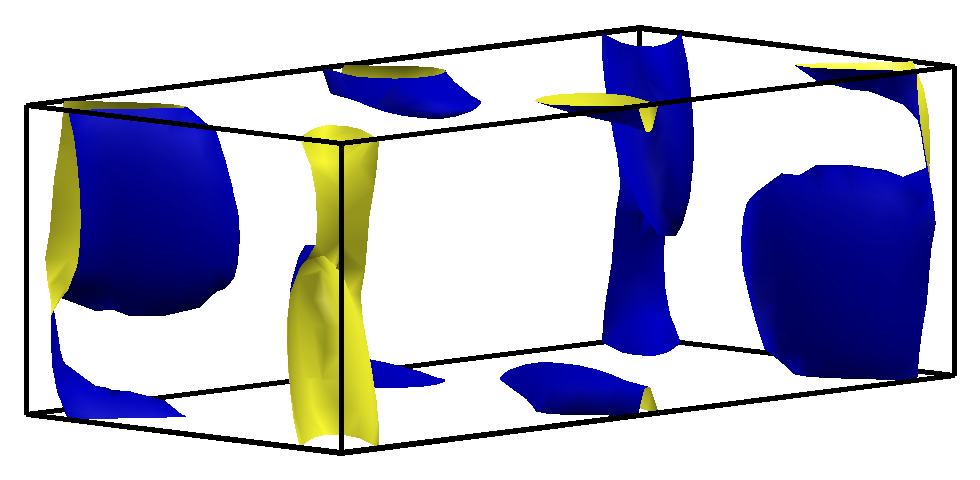}
  \includegraphics[width=0.4\textwidth]{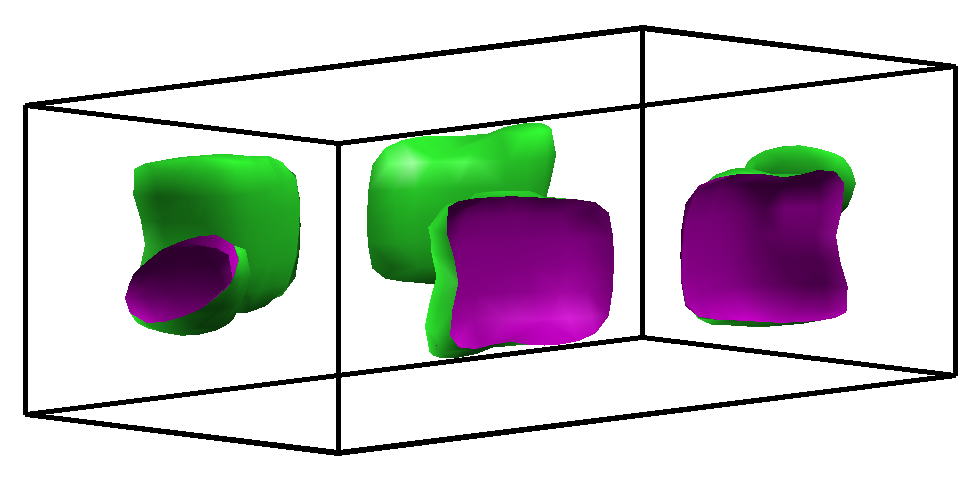}
  \qquad
  \includegraphics[width=0.4\textwidth]{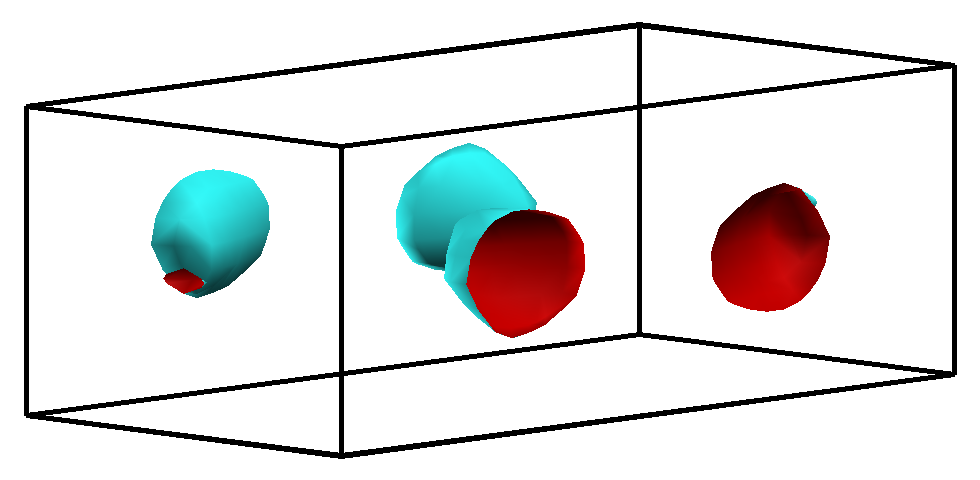}
  \caption{Left: Calculated Fermi sheets of LaNiO$_3$ in the
    disproportionated antiferromagnetic $P$-type phase. }
  \label{fig:ortho-br-fs}
\end{figure}

The Fermi surfaces of the disproportionated antiferromagnetic $R$-
and $P$-type phases, which are shown in Figs.~\ref{fig:rhomb-br-fs}
and \ref{fig:ortho-br-fs}, respectively, also illustrate this dramatic
change. It is remarkable that the band structure, DOS, and Fermi
surface change greatly due to the
$\uparrow\!\!0\!\!\downarrow\!\!0$-type antiferromagnetic ordering and
breathing distortions even though the total energies of the high- and
low-symmetry phases differ by only $\sim$2.0 meV/Ni. This indicates a
strong coupling between the electrons at the Fermi energy,
$\uparrow\!\!0\!\!\downarrow\!\!0$-type antiferromagnetic order, and
breathing distortions. The longitudinal magnetic fluctuations
associated with this coupling might damp the magnitude of the moments
at the Ni sites and may explain the small, so far undetermined value
of the ordered moments in LaNiO$_3$ \cite{Guo2018}.

The disproportionated antiferromagnetic phases of LaNiO$_3$ are
semimetallic and have band crossings generating small Fermi pockets.
In YNiO$_3$, however, a gap appears due to a complete splitting of the
lower half of the $e_g$-derived manifold. 
%

The Fermi pockets in the disproportionated antiferromagnetic phases
of LaNiO$_3$ are highly anisotropic. As can be seen in the right
panels of Figs.~\ref{fig:rhomb-bs} and \ref{fig:ortho-bs}, these
result from the Fermi level crossing of the bands that are highly
dispersive along certain directions and flat along
others. Furthermore, the band crossings occur near the edges and faces
of the Brillouin zone, which leads to the presence of valley
degeneracies. These characteristics cause the densities of states at
the Fermi level to be relatively large in the low-symmetry phases even
though the Fermi pockets enclose small portions of the Brillouin zone,
yielding small carrier concentrations. I obtain DOS values of 0.58 and
0.56 states eV$^{-1}$ per Ni per spin for the $R$-type and $P$-type
phases, respectively. This yields a calculated Sommerfeld coefficient
of $\gamma$ $\sim$2.7 mJ mol$^{-1}$ K$^{-2}$, which is a reduction of
around 40\% compared to the nondisproportionated, nonmagnetic phases.

The semimetallic electronic structure that I have obtained here for
the disproportionated antiferromagnetic phases agrees with the results
of Guo \textit{et al.}\ who find a metallic conductivity also in the
low-temperature antiferromagnetic phase of LaNiO$_3$
\cite{Guo2018}. In their electrical resistivity measurements, they
find that the resistivity of LaNiO$_3$ decreases even more rapidly
below the antiferromagnetic transition. This suggests a suppression of
the scattering channels below the transition temperature. Such a
behavior is also found in LaFeAsO, where the resistivity decreases at
a faster rate below the antiferromagnetic and structural phase
transition \cite{Klauss2008,McGuire2008}. What is incredible about
LaNiO$_3$ is that its resistivity is in the $\mu\Omega$ cm range in
Guo \textit{et al.}'s measurements. This is three orders of magnitude
lower than that of LaFeAsO, which has resistivity in the m$\Omega$ cm
range \cite{Klauss2008,McGuire2008}. The presence of highly dispersive
band crossings at the Fermi level in both the nondisproportionated,
nonmagnetic and disproportionated, antiferromagnetic phases might
underlie this behavior.

Another striking feature of Guo \textit{et al.}'s measurments is the
large value of 17 mJ mol$^{-1}$ K$^{-2}$ obtained for the Sommerfeld
coefficient $\gamma$, which is around five times larger than the value
calculated here for the disproportionated antiferromagnetic
phases. Previous ARPES
\cite{Eguchi2009,King2014,Yoo2015,Nowadnick2015}, optical conductivity
\cite{Ouellette2010,Stewart2011a,Stewart2011b}, and thermodynamics
measurements \cite{Zhou2014} have also identified large electron mass
enhancement and possible formation of pseudogapped state in LaNiO$_3$,
and this has been discussed in terms of strong local electronic
correlations. When the values for the on-site Coulomb $U$ and Hund's
rule $J$ are used such that they reproduce the experimentally measured
electronic structure, DMFT calculations can explain such an
enhancement \cite{Yoo2015,Nowadnick2015}. The calculations presented
here show that different structural phases occur close in energy in
both the nonmagnetic and antiferromagnetic LaNiO$_3$, and the nonlocal
fluctuations between these structures might provide an additional
avenue for the enhancement. The enhancement could also occur due to
the longitudinal magnetic fluctuations arising out of the strong
coupling between the electrons near the Fermi level, breathing
distortions, and {$\uparrow\!\!0\!\!\downarrow\!\!0$}
antiferromagnetic order. In particular, the pseudogap features
observed in LaNiO$_3$ are likely the result of the changes in the
electronic structure caused by the inchoate disproportionated
antiferromagnetism present in this material.

\section{Summary and Conclusions}
\label{sec:summ}

This work was motivated by three recent experimental studies on
LaNiO$_3$. i) The pair density analysis of the neutron scattering data
on a powder sample by Li \textit{et al.}, which showed that the
nanoscale structure of LaNiO$_3$ can be best described by the $Pnma$
and $P2_1/n$ structures above and below 200 K, respectively
\cite{Li2016}. ii) The x-ray diffraction, transport, and thermodynamic
experiments on single crystal samples by Zhang \textit{et al.}\ that
showed the material to be rhombohedral, metallic, and paramagnetic
down to 1.8 K \cite{Zhang2017}. iii) The neutron scattering,
transport, and thermodynamic experiments on single crystal samples by
Guo \textit{et al.}\ that showed an antiferromagnetic transition at 157
K but no structural and metal-insulator transitions \cite{Guo2018}.

I used DFT calculations to explore the structural, electronic, and
magnetic instabilities in LaNiO$_3$ indicated by these experiments.
The non-spin-polarized phonon dispersions of cubic LaNiO$_3$ show
instabilites at the wave vectors $R$ $(\ffrac12,\ffrac12,\ffrac12)_c$
and $M$ $(\ffrac12,\ffrac12,0)_c$ in the pseudocubic notation. I
relaxed different supercells with all possible Glazer tilts allowed by
these instabilities and found that several structures lie close in
energy. In my calculations, the $Pnma$ phase is marginally lower in
energy than the $R\overline{3}c$ phase. This suggests the presence of
structural fluctuations at finite temperatures and a possible
proximity to a structural quantum critical point. I was able to
stabilize several $\uparrow\uparrow\downarrow\downarrow$
antiferromagnetic configurations consistent with the experimentally
observed wave vector $(\ffrac14,\ffrac14,\ffrac14)_c$ in both
structural phases.  These occur in 20-atom $1\times1\times2$ and
80-atom $2\times2\times1$ supercells of the $R\overline{3}c$ and
$Pnma$ structures, respectively.  In both these structures, the
antiferromagnetic ordering caused an energy gain of only 0.7--0.4
meV/Ni relative to the respective nonmagnetic phases. The magnetic
moment per Ni in these configurations is 0.2 $\mu_B$.

The antiferromagnetic states are highly susceptible to the octahedral
breathing distortions with rock-salt ordering. Both phases relaxed to
the disproportionated $\uparrow\!\!0\!\!\downarrow\!\!0$ state with
moments of 0.6 $\mu_B$ and zero at the Ni sites inside the large and
small oxygen octahedra, respectively. The energies of both the
$R\overline{3}c$- and $Pnma$-derived disproportionated
antiferromagnetic phases are around 2.0 meV/Ni lower than the
respective nonmagnetic phases. The larger energy gain due to the
breathing distortions indicate that the disproportionation plays a key
role in the phase transition of the rare-earth nickelates, as
suggested by Mazin \textit{et al.}\ \cite{Mazin2007}. The appearance
of the breathing distortions in the $R\overline{3}c$-derived phase is
at variance with an earlier theoretical work, which suggested that the
magnetism in the $R\overline{3}c$ phase will occur without
disproportionation \cite{Lee2011a}.

The disproportionated antiferromagnetic phases derived from both the
$R\overline{3}c$ and $Pnma$ structures are semimetallic with small
three-dimensional Fermi pockets. This is consistent with the recent
results of Guo \textit{et al.}\ who observed an antiferromagnetic
transition in LaNiO$_3$ without a concomitant metal-insulator
transition \cite{Guo2018}. They did not observe the structural
distortions that I find, perhaps because the difference between the
Ni-O bond lengths in the large and small octahedra is only about
0.01--0.02 \AA. The transition to the disproportionated
antiferromagnetic phases causes a large change in the electronic
structure, and this may explain the observation of a pseudogap in the
earlier optical and tunneling spectroscopy experiments. In addition,
the structural and longitudinal magnetic fluctuations suggested by
these calculations may provide an explanation for the high electron
mass enhancement observed in this material.

\section*{Acknowledgements}

I am grateful to Beno\^it Fauqu\'e for helpful discussions. This work
was supported by the European Research Council grant ERC-319286 QMAC
and the Swiss National Supercomputing Center (CSCS) under project
s575.

\begin{appendix}

\section{Appendix}

\begin{figure}
  \centering
  \includegraphics[width=0.38\textwidth]{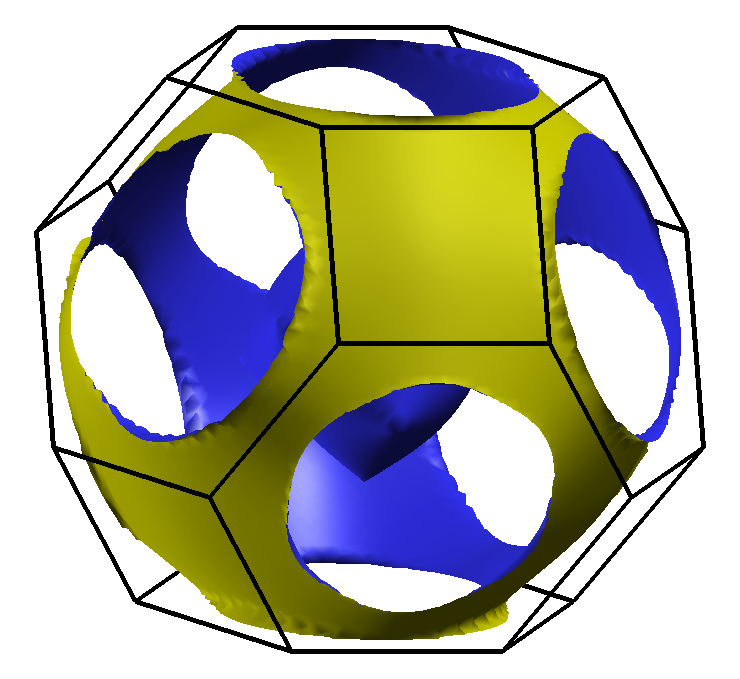}
  \qquad
  \includegraphics[width=0.38\textwidth]{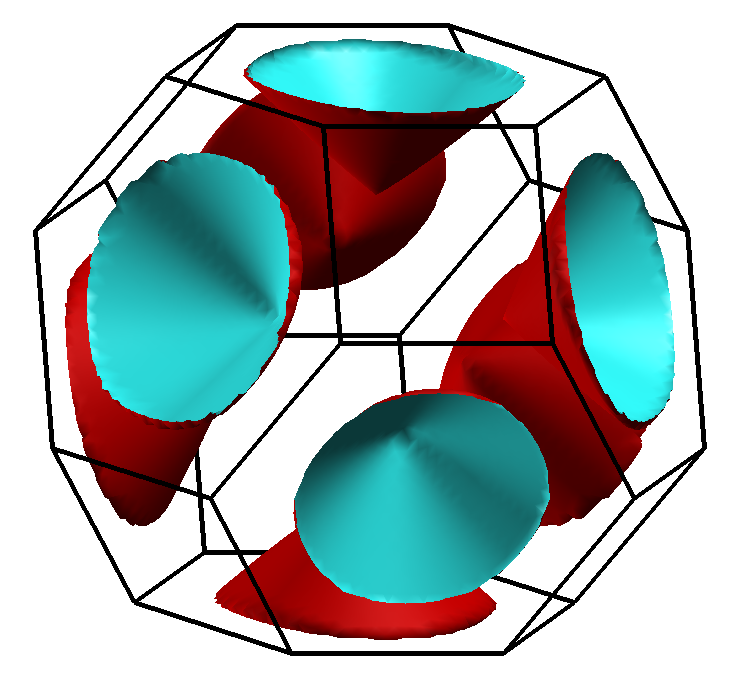}
  \caption{Calculated Fermi sheets of LaNiO$_3$ in the
    nondisproportionated, nonmagnetic $R\overline{3}c$ phase.}
  \label{fig:rhomb-fs}
\end{figure}

\begin{figure}[h]
  \centering
  \includegraphics[width=0.38\textwidth]{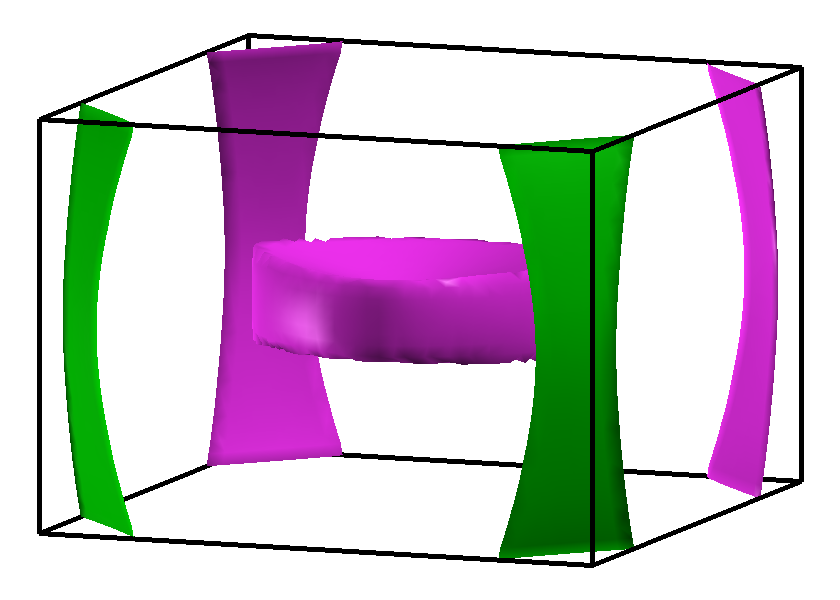}
  \qquad
  \includegraphics[width=0.38\textwidth]{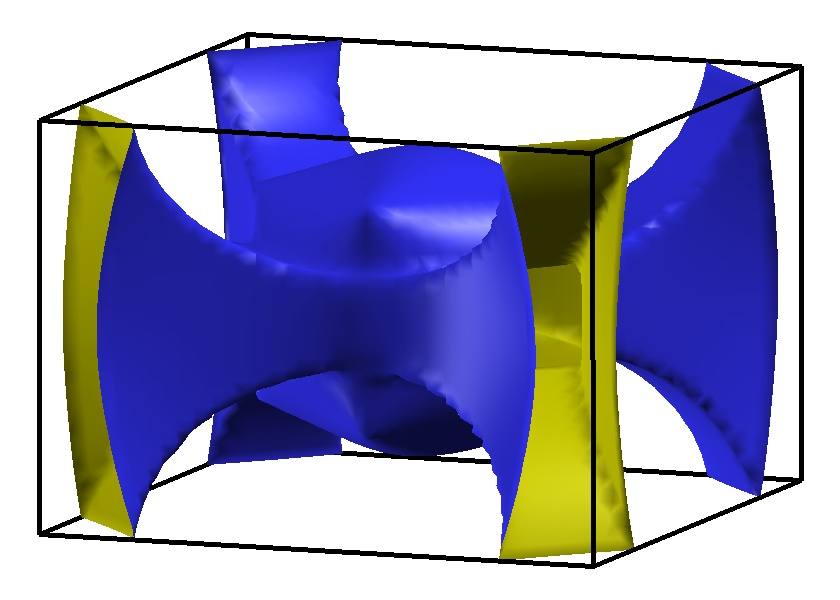}
  \includegraphics[width=0.38\textwidth]{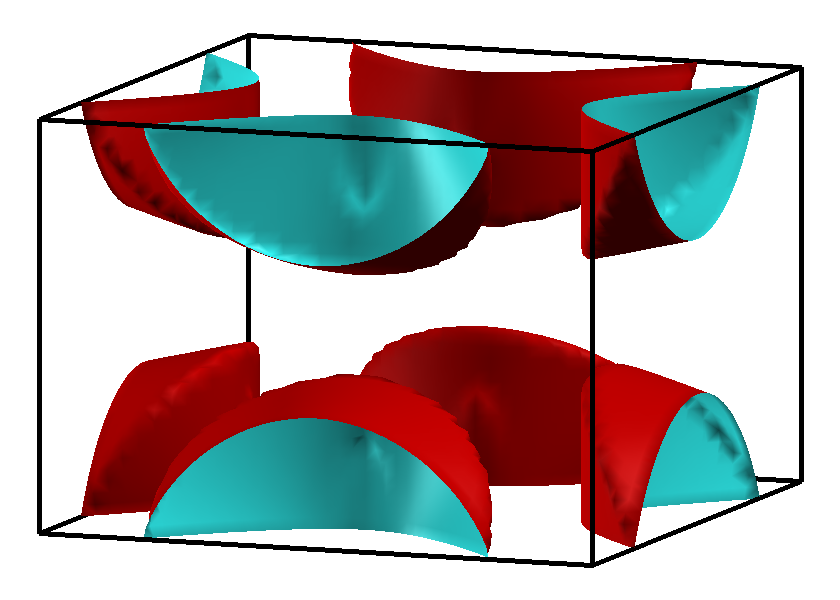}
  \qquad
  \includegraphics[width=0.38\textwidth]{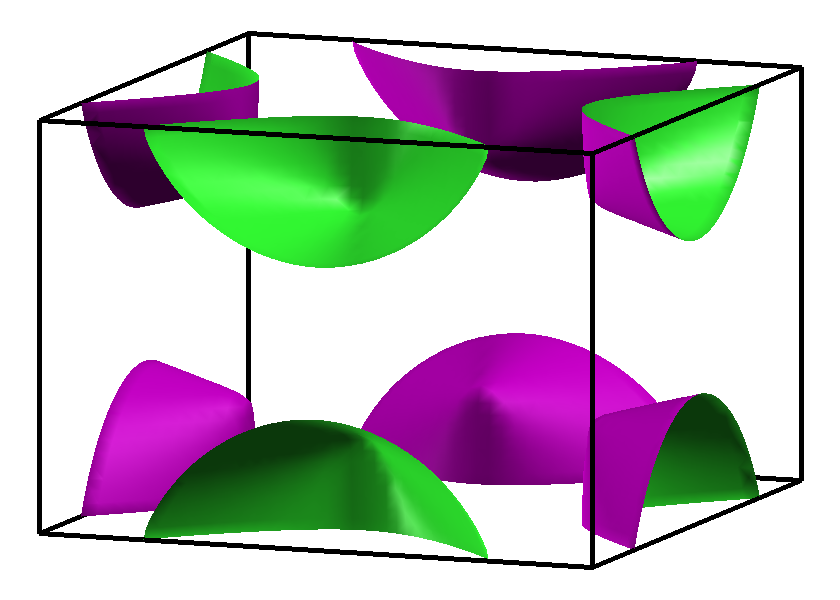}
  \caption{Calculated Fermi sheets of LaNiO$_3$ in the
    nondisproportionated, nonmagnetic $Pnma$ phase. }
  \label{fig:ortho-fs}
\end{figure}

For reference, the Fermi surfaces of the nondisproportionated,
nonmagnetic $R\overline{3}c$ and $Pnma$ phases are give in
Figs.~\ref{fig:rhomb-fs} and \ref{fig:ortho-fs}, respectively.

The relaxed atomic positions of the disproportionated
antiferromagnetic $R$ and $P$ phases are given in Tables
\ref{tab:rphase} and \ref{tab:pphase}, respectively. In the tables,
Ni$_{\uparrow}$, Ni$_{\downarrow}$, and Ni$_0$ denote Ni sites with
spins up, down, and zero, respectively.

\begin{table}
  \centering
  \caption{\label{tab:rphase} The relaxed atomic positions of the
    disproportionated antiferromagnetic $R$ phase. The atomic
    positions are given relative to the primitive lattice vectors. The
    lattice parameters are $a = 5.4064, b = 5.4078, c = 10.8092$ \AA,
    $\alpha = 61.306^\circ, \beta = 61.315^\circ,$ and $\gamma =
    61.167^\circ$. }
  \vspace{0.1in}
  \begin{tabular}{l d{1.4} d{1.4} d{1.4}}
    \toprule
    atom & \multicolumn{1}{c}{$x$} &
    \multicolumn{1}{c}{$y$} & \multicolumn{1}{c}{$z$}  \\
    \midrule
La   &    0.7502  &   0.7494  &    0.3751 \\
La   &    0.7502  &   0.7494  &    0.8751 \\
La   &    0.2498  &   0.2506  &    0.1249 \\
La   &    0.2498  &   0.2506  &    0.6249 \\
Ni$_{\uparrow}$    &    0.0000  &   0.0000  &    0.0000 \\
Ni$_{\downarrow}$  &    0.0000  &   0.0000  &    0.5000 \\
Ni$_{0}$  &    0.5000  &   0.5000  &    0.2500 \\
Ni$_{0}$  &    0.5000  &   0.5000  &    0.7500 \\
O       &    0.2486  &   0.8070  &    0.3455 \\
O       &    0.2486  &   0.8071  &    0.8454 \\
O       &    0.7514  &   0.1929  &    0.1545 \\
O       &    0.7514  &   0.1930  &    0.6545 \\
O       &    0.8042  &   0.6936  &    0.1253 \\
O       &    0.8042  &   0.6936  &    0.6253 \\
O       &    0.1958  &   0.3064  &    0.3747 \\
O       &    0.1958  &   0.3064  &    0.8747 \\
O       &    0.6912  &   0.2553  &    0.4033 \\
O       &    0.6912  &   0.2553  &    0.9033 \\
O       &    0.3088  &   0.7447  &    0.0967 \\
O       &    0.3088  &   0.7447  &    0.5967 \\
    \bottomrule
  \end{tabular}
\end{table}

\begin{table}
  \centering
  \caption{\label{tab:pphase} The relaxed atomic positions of the
    disproportionated antiferromagnetic $P$ phase. The atomic
    positions are given relative to the primitive lattice vectors. The
    lattice parameters are $a = 10.8835, b = 15.4253, c = 5.4979$ \AA,
    $\alpha = 90.028^\circ, \beta = 89.977^\circ$, and $\gamma =
    90.070^\circ$.}
  \vspace{0.1in}
  \begin{tabular}{l d{1.4} d{1.4} d{1.4} l d{1.4} d{1.4} d{1.4}}
    \toprule
    atom & \multicolumn{1}{c}{$x$} & \multicolumn{1}{c}{$y$} &
    \multicolumn{1}{c}{$z$} & atom & \multicolumn{1}{c}{$x$} & 
    \multicolumn{1}{c}{$y$} & \multicolumn{1}{c}{$z$}  \\
    \midrule
La          &    0.4880  &   0.3751   &   0.9955   &  O  &    0.2474   &   0.1244   &   0.9367   \\
La          &    0.4880  &   0.8751   &   0.9955   &  O  &    0.2474   &   0.6244   &   0.9367   \\
La          &    0.9880  &   0.3751   &   0.9955   &  O  &    0.7474   &   0.1244   &   0.9367   \\
La          &    0.9880  &   0.8751   &   0.9955   &  O  &    0.7474   &   0.6244   &   0.9367   \\
La          &    0.2381  &   0.3750   &   0.5044   &  O  &    0.4975   &   0.1256   &   0.5634   \\
La          &    0.2381  &   0.8750   &   0.5044   &  O  &    0.4975   &   0.6256   &   0.5634   \\
La          &    0.7381  &   0.3750   &   0.5044   &  O  &    0.9975   &   0.1256   &   0.5634   \\
La          &    0.7381  &   0.8750   &   0.5044   &  O  &    0.9975   &   0.6256   &   0.5634   \\
La          &    0.0120  &   0.1249   &   0.0045   &  O  &    0.3863   &   0.0171   &   0.2258   \\
La          &    0.0120  &   0.6249   &   0.0045   &  O  &    0.3863   &   0.5171   &   0.2258   \\
La          &    0.5120  &   0.1249   &   0.0045   &  O  &    0.8863   &   0.0171   &   0.2258   \\
La          &    0.5120  &   0.6249   &   0.0045   &  O  &    0.8863   &   0.5171   &   0.2258   \\
La          &    0.2620  &   0.1250   &   0.4956   &  O  &    0.1137   &   0.4829   &   0.7743   \\
La          &    0.2620  &   0.6250   &   0.4956   &  O  &    0.1137   &   0.9829   &   0.7743   \\
La          &    0.7620  &   0.1250   &   0.4956   &  O  &    0.6137   &   0.4829   &   0.7743   \\
La          &    0.7620  &   0.6250   &   0.4956   &  O  &    0.6137   &   0.9829   &   0.7743   \\
Ni$_{\uparrow}$   &    0.0000  &   0.0000   &   0.5000   &  O  &    0.3880   &   0.2329   &   0.2268   \\
Ni$_{\uparrow}$   &    0.5000  &   0.5000   &   0.5000   &  O  &    0.3880   &   0.7329   &   0.2268   \\
Ni$_{\uparrow}$   &    0.2500  &   0.2500   &   0.0000   &  O  &    0.8880   &   0.2329   &   0.2268   \\
Ni$_{\uparrow}$   &    0.7500  &   0.7500   &   0.0000   &  O  &    0.8880   &   0.7329   &   0.2268   \\
Ni$_{\downarrow}$   &    0.0000  &   0.4500   &   0.5000   &  O  &    0.1120   &   0.2671   &   0.7732   \\
Ni$_{\downarrow}$   &    0.5000  &   0.0000   &   0.5000   &  O  &    0.1120   &   0.7671   &   0.7732   \\
Ni$_{\downarrow}$   &    0.2500  &   0.7500   &   0.0000   &  O  &    0.6120   &   0.2671   &   0.7732   \\
Ni$_{\downarrow}$   &    0.7500  &   0.2500   &   0.0000   &  O  &    0.6120   &   0.7671   &   0.7732   \\
Ni$_0$      &    0.0000  &   0.2500   &   0.5000   &  O  &    0.3628   &   0.4827   &   0.7284   \\
Ni$_0$      &    0.5000  &   0.7500   &   0.5000   &  O  &    0.3628   &   0.9827   &   0.7284   \\
Ni$_0$      &    0.7500  &   0.0000   &   0.0000   &  O  &    0.8628   &   0.4827   &   0.7284   \\
Ni$_0$      &    0.2500  &   0.5000   &   0.0000   &  O  &    0.8628   &   0.9827   &   0.7284   \\
Ni$_0$      &    0.2500  &   0.0000   &   0.0000   &  O  &    0.1372   &   0.0173   &   0.2716   \\
Ni$_0$      &    0.7500  &   0.5000   &   0.0000   &  O  &    0.1372   &   0.5173   &   0.2716   \\
Ni$_0$      &    0.0000  &   0.7500   &   0.5000   &  O  &    0.6372   &   0.0173   &   0.2716   \\
Ni$_0$      &    0.5000  &   0.2500   &   0.5000   &  O  &    0.6372   &   0.5173   &   0.2716   \\
O      &    0.2526  &   0.3756   &   0.0633   &  O  &    0.3629   &   0.2671   &   0.7240   \\
O      &    0.2526  &   0.8756   &   0.0633   &  O  &    0.3629   &   0.7671   &   0.7240   \\
O      &    0.7526  &   0.3756   &   0.0633   &  O  &    0.8629   &   0.2671   &   0.7240   \\
O      &    0.7526  &   0.8756   &   0.0633   &  O  &    0.8629   &   0.7671   &   0.7240   \\
O      &    0.0025  &   0.3744   &   0.4366   &  O  &    0.1371   &   0.2329   &   0.2760   \\
O      &    0.0025  &   0.8744   &   0.4367   &  O  &    0.1371   &   0.7329   &   0.2760   \\
O      &    0.5025  &   0.3744   &   0.4366   &  O  &    0.6371   &   0.2329   &   0.2760   \\
O      &    0.5025  &   0.8744   &   0.4366   &  O  &    0.6371   &   0.7329   &   0.2760   \\
    \bottomrule
  \end{tabular}
\end{table}

%
%
\end{appendix}

\bibliography{lno3-paper}

\nolinenumbers

\end{document}